\documentclass[twocolumn]{aastex63}

\usepackage{graphicx}
\usepackage{float}

\usepackage{aas_macros}
\usepackage{color}
\usepackage{url}

\usepackage{enumitem}
\usepackage{booktabs}
\usepackage{multirow}
\newcommand{\red}{\textcolor{red}}

\shorttitle{Kinematically-persistent satellite planes}  
\shortauthors{Santos-Santos et al.}

\begin{document}
 
\title{
Planes of satellites around simulated disk galaxies II: Time-persistent planes of kinematically-coherent satellites in $\Lambda$CDM}

\author{Isabel Santos-Santos}
\affiliation{Institute for Computational Cosmology, Department of Physics, Durham University, South Road, Durham, DH1 3LE, UK}

\author{Mat\'ias G\'amez-Mar\'in}
\affiliation{Departamento de F\'isica Te\'orica, Universidad Aut\'onoma de Madrid, E-28049 Cantoblanco, Madrid, Spain}

\author{Rosa Dom\'inguez-Tenreiro}
\affiliation{Departamento de F\'isica Te\'orica, Universidad Aut\'onoma de Madrid, E-28049 Cantoblanco, Madrid, Spain}
\affiliation{Centro de Investigaci\'on Avanzada en F\'isica Fundamental, Universidad Aut\'onoma de Madrid, E-28049 Cantoblanco, Madrid, Spain}

\author{Patricia B. Tissera}
\affiliation{Instituto de Astrof\'isica, Facultad de F\'isica, Pontificia Universidad Cat\'olica de Chile, Av.Vicu\~na Mackenna 4860, Santiago, Chile}
\affiliation{Centro de Astro-Ingenier\'ia, Pontificia Universidad Cat\'olica de Chile, Av. Vicu\~na Mackenna 4860, 7820436 Macul, Santiago, Chile}

\author{Lucas Bignone}
\affiliation{Instituto de Astronom\'ia y F\'isica del Espacio, CONICET-UBA, 1428, Buenos Aires, Argentina}

\author{Susana E. Pedrosa}
\affiliation{Instituto de Astronom\'ia y F\'isica del Espacio, CONICET-UBA, 1428, Buenos Aires, Argentina}

\author{H\'ector Artal}
\affiliation{Tecplot, Inc., Rutherford, NJ 07070, USA}

\author{M.\'Angeles G\'omez-Flechoso}
\affiliation{Departamento de F\'isica de la Tierra y Astrof\'isica, Universidad Complutense de Madrid, E-28040 Madrid, Spain}
\affiliation{Instituto de F\'isica de Part\'iculas y del Cosmos (IPARCOS) ,Universidad Complutense de Madrid, E-28040 Madrid, Spain}

\author{V\'ictor Rufo-Pastor}
\affiliation{Departamento de F\'isica Te\'orica, Universidad Aut\'onoma de Madrid, E-28049 Cantoblanco, Madrid, Spain}

\author{Francisco Mart\'inez-Serrano}
\affiliation{Dassault Systemes S.E.,  28020 Madrid, Spain}

\author{Arturo Serna}
\affiliation{Departamento de F\'isica Aplicada, Univ. Miguel Hern\'andez, Elche, Spain}

\begin{abstract}
We use two zoom-in $\Lambda$CDM hydrodynamical simulations of massive disk galaxies to study 
the possible existence
of fixed satellite groups showing a kinematically-coherent behaviour across evolution (angular momentum conservation and clustering). We identify three such groups in the two simulations, defining kinematically-coherent, time-persistent planes (KPPs) that last 
at least from virialization to $z=0$ (more than 7 Gyrs).
This proves that orbital pole clustering is not necessarily set in at low redshift, representing a long-lived property of galaxy systems. KPPs are thin and oblate, represent $\sim25-40\%$ of the total number of satellites in the system, and are roughly perpendicular to their corresponding central disk galaxies during certain periods, consistently with Milky Way $z=0$ data. KPP satellite members are statistically distinguishable from satellites outside KPPs: they show higher specific orbital angular momenta, orbit more perpendicularly to the central disk galaxy, and have larger pericentric distances, than the latter. We numerically prove, for the first time, that KPPs and the best-quality positional planes share the same space configuration 
across time,
such that KPPs act as `skeletons' preventing the latter of being washed out in short timescales. In one of the 
satellite-host systems, we witness the late capture of a massive dwarf galaxy endowed with its own satellite system, also organized into a KPP configuration prior to its capture. We briefly explore the consequences this event has on the host's KPP, and on the possible enhancement of the asymmetry in the number of  
satellites rotating in one sense or the opposite within the KPP.
\end{abstract}



\section{Introduction}\label{sec:intro}
Planar alignments of satellites observed in the local Universe have been considered for a long time as one of the most challenging small-scale issues for $\Lambda$CDM.

Recently, the obtention of proper motion data for MW satellites (specially  by the \textit{Gaia} mission) has allowed to compute their 3D velocities and orbital angular momentum vectors, and study the  plane of satellites in the MW as a kinematic structure.
Disentangling  the  clues on  the formation of this structure and its stability, in terms of basic physical laws, seems now a more achievable goal than    just from  spatial data on satellite positions.

The study of positional satellite data alone
raised the problem of satellite planes.
Later,
 with the help of numerical simulations,  it was possible to evaluate their high significance.
Indeed,  it was long-ago noticed that the `classical' (11 brightest)
 satellites  were arranged in a common planar structure that is approximately perpendicular to the Galactic disk
 \citep{Lynden76,Kunkel76,Kroupa05}.
In the past decade it has been shown that all known MW satellites   fall  on the same planar structure, i.e., the so-called  `VPOS' for `vast polar structure' \citep{Pawlowski12,Pawlowski13}.
New updatings\footnote{We caution the reader about possible biases as  the sky has not been homogeneously scanned in search of MW satellites. Most of the newly-added ultrafaints are in the LMC region \citep[e.g.][]{Drlica-Wagner2015}.}, including ultrafaint satellites, indicate that these add to the VPOS \citep[][ hereafter Paper I]{SantosSantos2020I}.
M31's satellites
 are also anisotropically distributed \citep{Koch06,Metz07,McConnachie06}, with half of the satellite population
forming a thin planar structure, referred to as the `GPoA', or `Great Plane of Andromeda', \citep{Conn13,Ibata13,Pawlowski13,SantosSantos2020I}.
In addition, a second positional plane of satellites, perpendicular to the former  and with comparable population and thickness, was  identified in Paper I.

Flattened satellite distributions have also been observed beyond the MW and M31, the best analyzed system being Cen A, a massive galaxy with clear signals of a recent merger, see e.g. \citet{Tully2015}, \citet{Muller2016} and \citet{Muller2021}. In their respective works, \citet{Chiboucas2013}, \citet{Muller2017} and \citet{MtnezDelgado2021} discuss the possibility of flattened satellite distributions around M81, M101 and the NGC 253 galaxies.

While positional data opened very interesting debates on the issue of satellite planes, the  availability of proper motions has  opened the possibility of studying them as kinematic entities.

The MW is the only system with 3D kinematic data for a large enough  satellite sample that studying these planes as kinematic structures  is warranted.
Results show that a high fraction of MW satellites present well-aligned orbital poles mostly perpendicular to the Galactic disk axis \citep{Metz08,Pawlowski13b,GaiaHelmi18,Fritz18,Pawlowski2020}.
Paper I studied the co-orbitation of MW satellites 
for which kinematic data were available,
 corresponding to 36 satellites as of August 2020.
Here, as in Paper I, the term  `co-orbitation' will mean kinematic coherence no matter the sense of rotation within an aperture $\alpha_{\rm co-orbit} = 36.87^\circ$, see \citet{Fritz18}. 
When no distinction is made between one sense of rotation and the contrary within the kinematic structure, it was found that a fraction  between $\sim 25-50\%$ of satellites (taking into account the proper motion uncertainties) show orbital poles within an area of $10\%$ of the sphere around the normal direction to the VPOS.
When a distinction is made between co- and counter-rotation within the kinematic plane, it was found that the mean fraction of co-rotating satellites  is $\sim 33\%$. From a total of 36 satellites, this corresponds to 12 satellites, with only 4 satellites in counter-rotation, i.e., a mean co- over counter-rotating ratio of 3/1
(and minimum ratio of $\sim5/3$ considering errors, see figure 11 in Paper I).

While little proper motion data are currently available for M31 satellites, line-of-sight velocities can be used to elucidate the co-orbitation of M31 satellites forming planes.
\citet{Ibata13} claim a coherent rotational motion of satellites on the GPoA plane seen edge-on from the MW, with the northern satellites receding and the southern ones approaching \citep[see also][]{Sohn2020}. On the other hand, line-of-sight velocities allow to calculate the perpendicular velocity dispersion  of the second positional plane of satellites identified in M31, as it is face-on from the MW: in Paper I it is found to be of $\sim 90$ km/s.
According to models by \citet{Fernando17}, planes with a perpendicular velocity dispersion above $\sim50$ km/s dissolve, ending up with half their initial number of satellites in 2 Gyrs time. Thus, M31's second plane would not be a stable structure.
These results however should be taken with caution, as distance data for M31 satellites are still  uncertain and method-dependent.
Therefore the existence of a second positional plane in M31 has still to be confirmed.

Similarly, only line-of-sight velocities are available for the Cen A system.
\citet{Muller2018,Muller2021} discuss the coherent structure of these velocity fields, possibly implying a rotating plane of satellites seen  edge-on.




There have been numerous theoretical studies within the $\Lambda$CDM cosmological framework addressing the frequency and origin of positionally-detected planar alignments of satellites like those observed. These have made use of cosmological simulations,  either `dark matter only (DMO)' \citep{Libeskind05,Libeskind09,Lovell2011,Wang13,Bahl14,Ibata2014,Buck15,Buck16,Cautun15},
  or including hydrodynamics \citep{Gillet15,Ahmed17,Maji17b,Garaldi18,Shao19,Samuel2021,Sawala2022}.
Positionally-detected planes of satellites as thin as the classical plane in the MW, or the GPoA plane in M31,  can be found in $\Lambda$CDM but are rare.
For example, in their analysis of the Illustris simulations, 
\citet{ForeroRomero2018}
focus on satellite systems within Local Group-like analogs, finding an important dependence on the particular properties of the systems and their evolution, with only $4\%$ of systems similar to the MW one.




Some studies have also investigated the kinematic coherence of satellite planes.
\citet{Gillet15} and \citet{Buck16}
found that good-quality, positionally-detected planes at $z=0$ are not fully kinematically-coherent, but only $\sim30-60\%$ of their members are.
\citet{Ahmed17} found consistent results  in their analysis of four 
$z=0$ halos where thin positional planes were detected: only two of these planes had satellite members whose orbital poles clustered
(either on a side or the opposite in a spherical projection, meaning they co-orbit).
\citet{SantosSantos2020II}, hereafter Paper II,   searched for positional planes of satellites in two different zoom-in hydro-simulations along cosmic evolution, using a method based on the `4-galaxy-normal' density plot \citep{Pawlowski13}. 
They  found, at each timestep,  planes as thin and thinner than  those observed, whose quality (measured by thinness and population) changes with time.
They also found that, in general, the number of co-orbiting satellites is just a fraction of the total number of satellites in the plane, fraction  changing abruptly with time as well.
In agreement with previous work, Paper II results suggested that high-quality positionally-detected planes are   unstable, transient structures,
probably due to the presence of an important fraction of satellites whose  membership is short-lived. These fortuitous interlopers  leave or join the structure on short timescales, giving rise to important fluctuations in the positional-plane properties as a function of time.

Instead of focusing on co-orbitation within  planes identified in positions, certain works have  directly analyzed satellite clustering in angular momentum space.
\citet{Garaldi18} studied 4 galactic systems, finding groups of satellites with clustered orbital angular momenta in two of them. These groups 
conformed up to $\sim 50\%$ of the total number of satellites. Interestingly, these two systems
have well-defined central \textit{disk} galaxies. In addition, both satellite groups define thin planes, but the kinematic coherence of satellites sets in at early times in one system
(where the ratio of satellites orbiting in one sense over those rotating in the opposite  is $\sim$7/3) and at late times in the other  (where all satellites rotate in the same sense).
Other studies have  tried to recover the specific kinematical coherence of the MW's `classical' satellites, where 8 out of 11 satellites co-rotate. \citet{Shao19} find that only $\sim 30\%$ of all the thin planes with 11 members in the EAGLE volume show such a coherence, and that orbital pole clustering is tighter on average at low redshift, suggesting that kinematic structures like that of the MW may be dynamically young and formed recently.
\citet{Sawala2022}  reached similar conclusions on the instability of  the MW's plane from satellite orbit integrations using recent proper motion data \citep[see also][]{Maji17b,Lipnicky17}.
Finally, a new ingredient in the issue of  kinematic-coherence is introduced by \citet{Samuel2021} who propound that the effect of an LMC-like galaxy on the satellite system favours the thinness and coherence of planes \citep[see also][]{GaravitoCamargo2021}. However, they also claim that the observed MW plane is likely to be a temporary configuration that will dissapear as soon as the LMC moves far away on its orbit\footnote{In their analysis of the DMO Millennium-II simulation, \citet{Ibata2014} had already claimed  that the rare ($0.04\%$) systems similar to  M31 they found include the infall of  a massive halo  carrying its own set  of satellites.}.

Finally, other studies focus on external systems beyond the Local Group. In their search of Cen A analogues in the Illustris-TNG simulation, 
\citet{Muller2021} find 0.2$\%$ of them. However, these are not co-orbiting systems, but short-lived, chance projections along the line of sight.

From this brief summary about previous results on satellite kinematical-coherence, we see that there is a lively debate on this issue, where most authors so far have found that positionally-detected planes  in $\Lambda$CDM simulations do not fully  consist of co-orbiting satellites, but only $\sim$half of them are: 
i.e.,
positional planes could consist of a fraction  of coherently-rotating satellites,
plus others that have unclustered orbital poles and that,  fortuitously, happen to be plane members  for a short period.
However, previous results 
are inconclusive
 with regards to
the role that kinematic  support can play in ensuring the stability and long-term duration of such planes,
the age at which a system acquired its fraction of co-orbitation  coherence,
and whether or not it persists once acquired.

Indeed, the  existence in a system of a subset of satellites   that mantains a coherent kinematical behaviour over a long period of time, is yet to be proven.
Should this happen, this subset of coherently-orbiting satellites would contribute a kind of skeleton ensuring a long-term durability to positional planes \citep[as suggested by][]{Gillet15,Buck16}.
In this case,
transient, non-kinematically-coherent satellites would be positional plane members only while they happen to cross it; after,  
they would be lost to the planar configuration and replaced by  other transient members.

Finally,  
it also remains unclear if there is need for some extra ingredient favouring plane kinematic and/or  rotational coherence, e.g., enhancing a  higher  fraction of co-orbiting  and/or co-rotating   orbits in the same sense.



In conclusion,
a  step forward in understanding the issue of the origin and possible stability and endurance  of satellite planes is their study as kinematic configurations.
By extending the analysis to the six-dimensional phase space, we can explore  the conservation of satellite orbital angular momentum over long time intervals:
if a fraction of their orbital pole directions happen to be clustered, and remain so, 
this could be a key to finding an answer. This is the aim of this paper.
We note that satellite orbital angular momentum conservation  cannot be taken for granted, as the gravitational field satellites feel is a changing one, suffering the effects of mass accretions onto halos and disks, as well as  interactions. Analyzing cosmological simulations is an adequate procedure to elucidate this question. 


To address this issue in a systematic and detailed manner, in this paper
we study the same simulated 
galactic systems
analyzed in Paper II throughout their evolution.
We analyze them paying particular attention to the issues related to satellite orbital angular momentum conservation and the clustering of their orbital pole directions, and their  consequences.
In this way we identify fixed sets of satellites whose orbital angular momenta are conserved over long time intervals and whose orbital poles remain clustered, giving rise to thin, persistent-in-time planes.
We shall refer to these planes as Kinematically-coherent Persistent Planes, or KPPs. 
We investigate how kinematically-identified plane properties evolve with time and quantify their persistence.
We also address in this paper  the possibility of  a second channel that could enhance 
kinematic-coherence, and, more specifically,  the fraction of  satellites co-rotating in the same sense,
involving the late capture of a massive dwarf endowed with its own (sub)-satellite system.

Satellites on a persistent plane suggest a scenario where
they could have gained their common dynamics at high redshift, probably in unison with the local configuration of the large-scale structure they are embedded in.
This issue will be addressed in a forthcoming paper (G\'amez-Mar\'{i}n  et al., in prep.).


The paper is organized as follows.
Sections~\ref{sims} and \ref{sec:satsamples} introduce the simulations analyzed and their corresponding satellite samples, respectively,
focusing on satellite orbit characteristics.
Section~\ref{sec:results} is devoted to the identification of stable-in-time axes of maximum satellite co-orbitation, and we present our results on satellite KPP  members  as well. 
In Section~\ref{sec:KineSat} the properties of KPPs as positional planes are studied.
In Section \ref{sec:PropKineSat} we analyze whether satellites in KPPs  have  properties making them a group of satellites that are statistically distinguisable from non-KPP ones.
Section \ref{sec:late-capture}  is devoted to a brief analysis of the late capture of a  massive dwarf carrying its own satellite system.
The relationships between KPPs and the best quality positional planes at fixed satellite number are discussed in
Section \ref{sec:KPPvsPosPla}. 
Finally, results are summarized in
 Section~\ref{sec:conclu}, where we also  expose the conclusions reached.


\section{Simulations}\label{sims}
We study planes of satellites orbiting around simulated isolated massive disk galaxies. 
The two simulations used in this work are Aq-C$^\alpha$  and PDEVA-5004.
They make use of different initial conditions, codes, and subgrid prescriptions, which will thus  grant our conclusions 
independency  on the many details of simulation modelling.

The simulations have been chosen based on the following criteria or characteristics:
(a) the galactic system contains a central galaxy with a thin stellar and gaseous disk at $z\sim0$ that presents a large radial extent ($\sim20$ kpc);
(b) the central galaxy shows an overall quiet merger history, especially free of major-merger events at late times, i.e., after virialization; 
(c) the system hosts a numerous ($\sim30$) satellite population around the central galaxy;
(d) the simulation presents a high enough mass resolution to allow for a minimum of 50 baryonic particles per satellite,
 ($M_{\rm bar}\sim10^7$M$_\odot$). This is in order to ensure that satellite centers of mass and velocity are computed with sufficient accuracy to analyze angular momentum conservation, see Section~\ref{sec:jorbevo}.

After a pre-analysis of a set of different zoom-in cosmological hydro-simulations we identified two that met the previous prerequisites. Details of each simulation are explained below.

\subsection{Aq-C$^\alpha$}
The initial conditions of this simulation come from the   Aquarius Project  \citep{Springel08}, a selection of DMO  MW-sized halos, formed in a $\Lambda$CDM 
$100 h^{-1}\, \rm Mpc$ side cosmological box.
In this work we analyze a new re-simulation of the so-called ``Aquarius-C" halo (hereafter Aq-C$^{\alpha}$),  including the hydrodynamic  and subgrid models  described in
\citet{Scannapieco05,Scannapieco06} with the modification presented in  \citet{Pedrosa15}.  
The initial mass resolution of baryonic and dark matter particles is 
$m_{\rm bar}=4.1\times10^5 \,\rm M_{\odot}$, and $m_{\rm dm}=2.2\times10^6 \,\rm M_{\odot}$, respectively,  
with a cosmological model characterized by
$\Omega_m$= 0.25; $\Omega_b$= 0.04; $\Omega_\Lambda$= 0.75; $\sigma_8$=0.9; $n_s$=1;
  $H_0$ =  73 $ \rm km s^{-1}\,Mpc^{-1}$.

A  standard two-phase process characterizes the halo mass assembly: first a fast  phase  where mass growth rates are high, mainly through merger activity,
and then a slow phase where  they decrease.
The  halo collapse or virialization,  i.e., the time when it gets decoupled from global expansion,
happens at  a Universe age of $T_{\rm vir,AqC} \simeq$ 7 Gyr ($z_{\rm vir}\simeq 0.76$). This is an important timescale in halo mass assembly history.
In this case, a 25\% of the mass is accreted after collapse.

This galaxy presents a quiet history from $z\approx1.5$ to $z\approx0.18$, where no major mergers occur.
Soonly after, the main galaxy undergoes a potentially disturbing dynamical interaction, as it captures a massive dwarf
(hereafter MD, $M_{\rm bar}\sim5\times10^{9}\,\rm M_\odot$), carrying its own satellite system.
We
 use this event in Aq-C$^\alpha$ to study its dynamical effects  on the original satellite system. To this end, we analyse the simulation 
 both neglecting and including the massive dwarf and its satellites. 

Properties of the main galaxy measured at redshift $z=0.18$ are:
 $M_\star=7.6\times10^{10}\,\rm M_\odot$, $M_{\rm gas}=5.6\times10^{10}\,\rm M_\odot$,
$M_{\rm vir}=1.5\times10^{12}\,\rm M_\odot$ and $R_{\rm vir}=219 \,\rm kpc$.

\subsection{PDEVA-5004}
The PDEVA-5004 system comes from a zoom-in re-simulation run with the PDEVA code \citep{MartinezSerrano08}
of a halo identified in a $\Lambda$CDM 10 Mpc per side periodic box, where the following parameters are assumed:
$\Omega_{\Lambda}$ = 0.723, $\Omega_{m}$ = 0.277, $\Omega_{b}$ = 0.04, and $h$ = 0.7.
The  PDEVA-5004 system \citep{domenech12}  has been  previously studied in several projects,
where satisfactory consistency with observational data has been found in all the comparisons addressed, see Paper II and references therein.
The mass resolution of baryonic and dark matter particles  is 
$m_{\rm bar}=3.9\times10^5 \, \rm M_{\odot}$, and    $m_{\rm dm}=1.9\times10^6\, \rm M_{\odot}$, 
respectively.

The halo growth history shows again a two-phase process with $T_{\rm vir,5004} \simeq 6$ Gyr. 
Only a 20\% of the virial mass is assembled after this time, and no major mergers occur.
At redshift $z=0$, PDEVA-5004's main galaxy presents the following properties: 
$M_\star=3.1\times10^{10}\,\rm M_\odot$, $M_{\rm gas}=8.6\times10^{9}\, \rm M_\odot$, $M_{\rm vir}=3.4\times10^{11}\, \rm M_\odot$, $R_{\rm vir}\approx 185$ kpc.
It is roughly less  massive and smaller than Aq-C$^\alpha$.

\section{Satellite samples}\label{sec:satsamples}
\begin{figure*}
\centering

\includegraphics[width=0.49\linewidth]{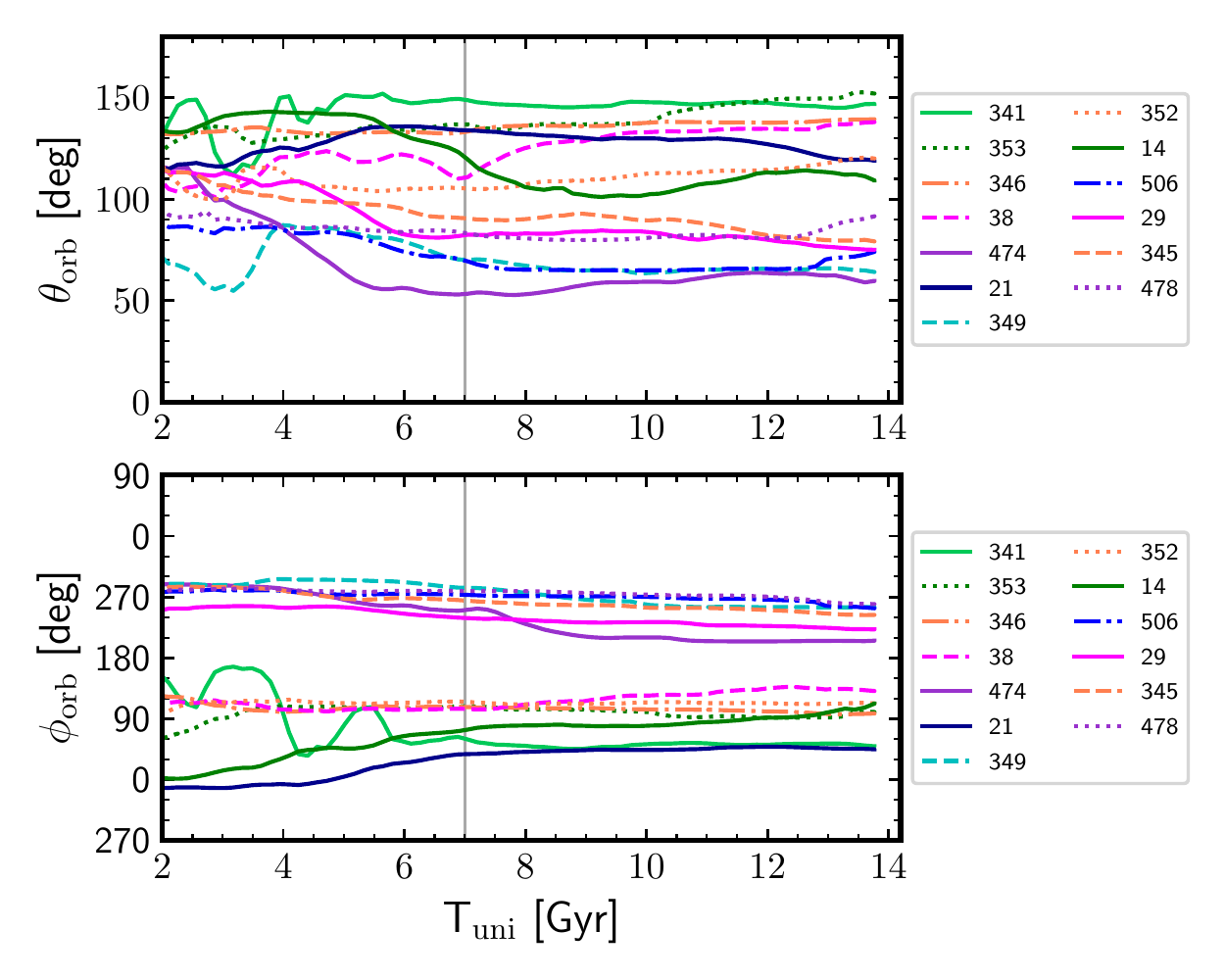}
\includegraphics[width=0.49\linewidth]{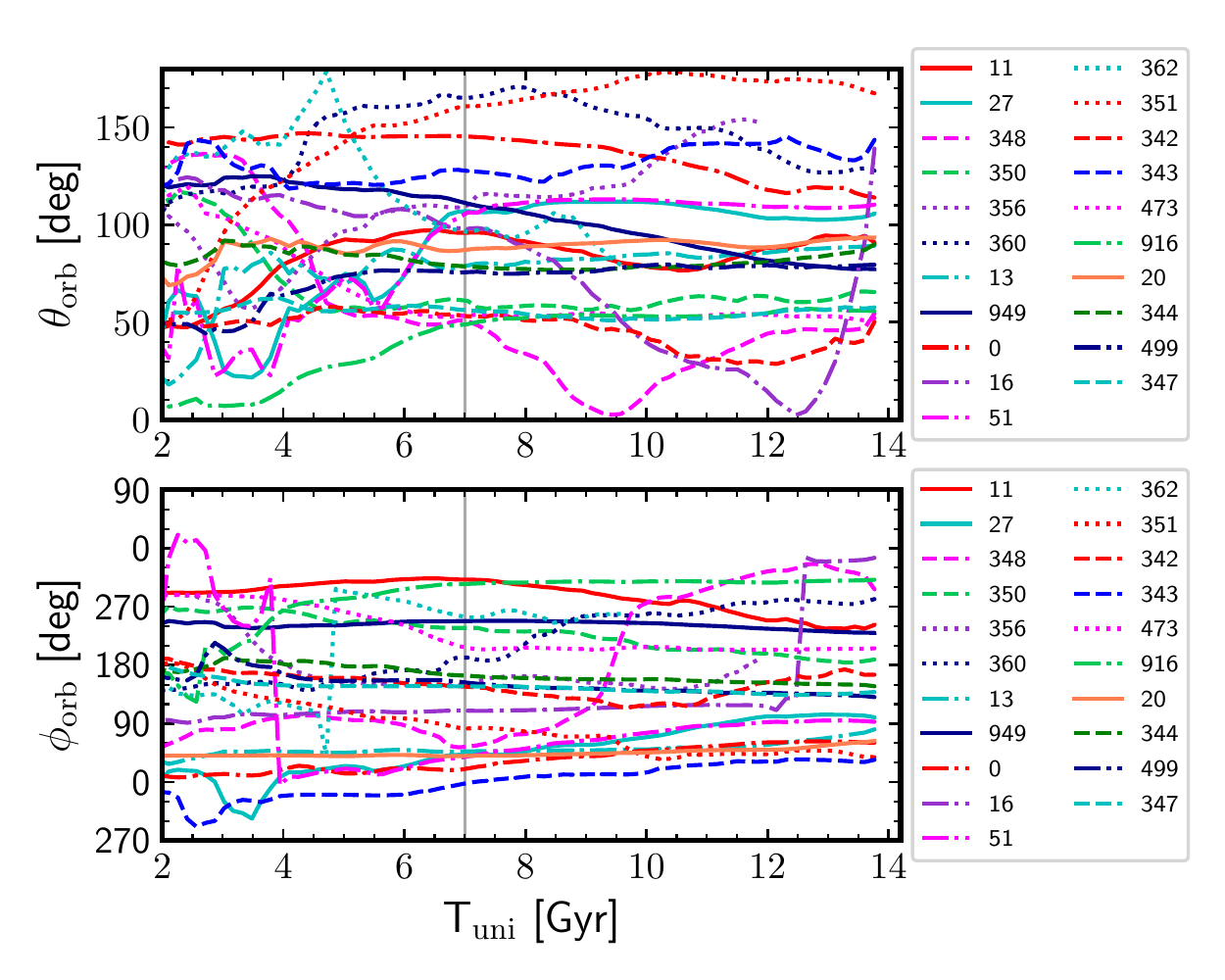}
\caption{
The angular components of the $\vec{J}_{orb}$ vector of each satellite, as a function of time, for satellites in the Aq-C$^{\alpha}$ simulation. Satellite samples are divided according to belonging (left panels) or not (right panels) to  kinematically-coherent, persistent planes, see Table \ref{table:summary-discard}, second column.
Vertical lines mark the halo virialization time, T$_{\rm vir}$, i.e., roughly the moment when halo decouples from cosmic expansion. Satellite IDs are color-coded in the legends.}
\label{fig:jsat_AqC}
\end{figure*}

\begin{figure*}
\centering
\includegraphics[width=0.8\linewidth]{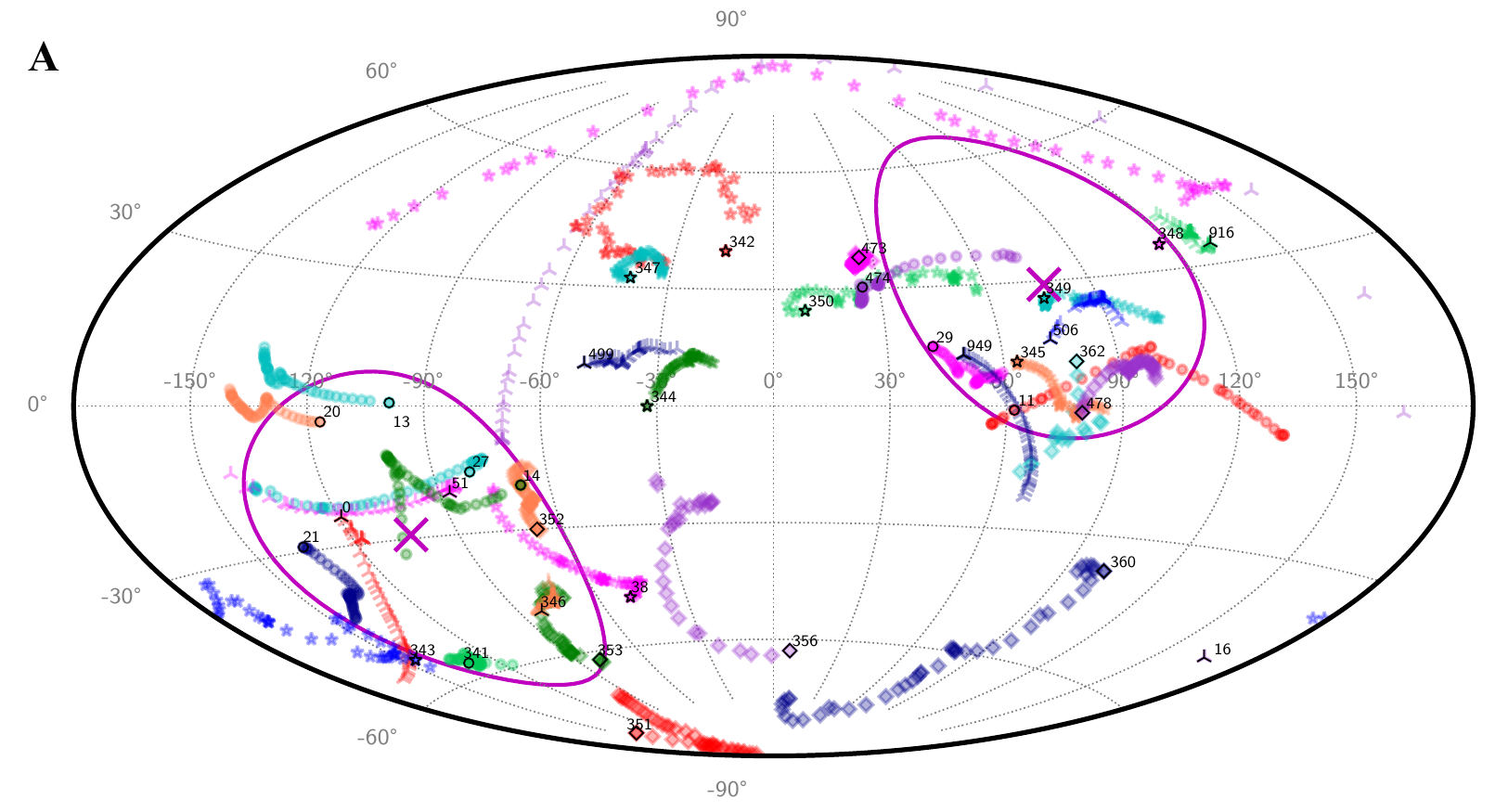}\\
\includegraphics[width=\linewidth]{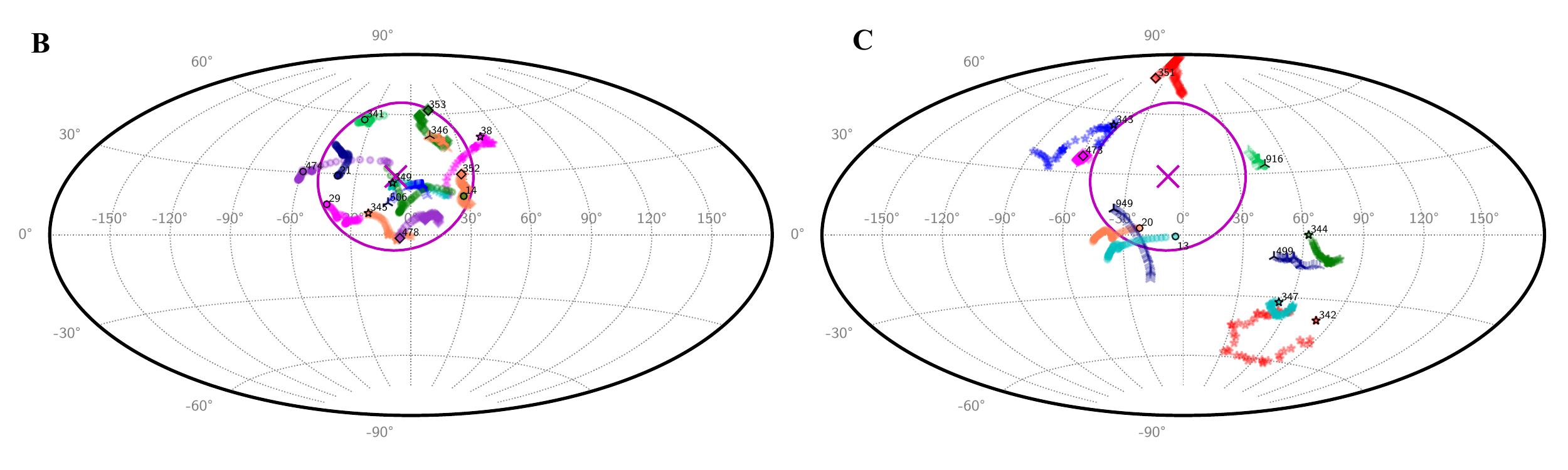}\\
\hspace{1.5cm}\includegraphics[width=0.55\linewidth]{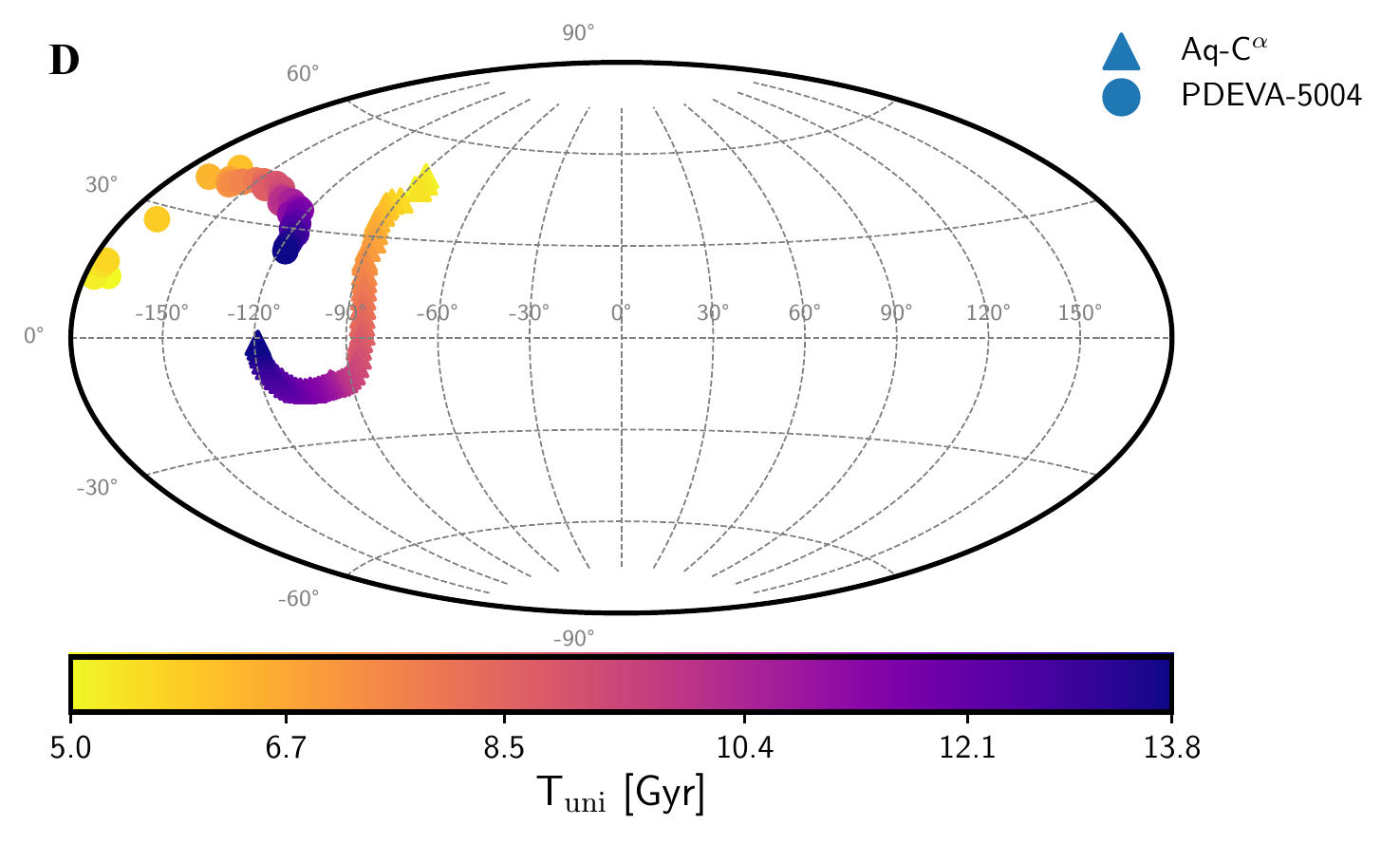}
\caption{Aitoff projections.
\textit{Panel A:} Orbital poles (i.e., the  $\vec{J}_{\rm orb}$ directions) for the whole   Aq-C$^\alpha$ satellite sample.  
Satellite identities  are specified through  point type and color, and different points stand for different output times from  T$_{\rm vir}$ to T$_{\rm z=0}$.
 Satellite ID  numbers are given next to their trace-point corresponding to $z=0$. 
`X' crosses mark the (axial) axis of maximum satellite co-orbitation, $\vec{J}_{\rm stack}$.
These crosses  are diametrically opposite.
The circle  around each of them  encloses the co-orbiting directions, i.e., those directions at angular distances  closer than  $\alpha_{\rm co-orbit}=36.87^\circ$ from each axis.
Satellites falling within these circles orbit on a common orbital plane, rotating in one sense or in the contrary.
\textit{Panels B \& C:} Results of KPP satellite membership determination  by means of the $stacking$  $method$.
In these plots  only the poles relative to those satellites involved in the stacking procedure are drawn (i.e., those with well-conserved orbital angular momentum, see Section~\ref{sec:methodsims}).
\textit{Panel B:} Satellites belonging to  the KPP  (second  column of Table \ref{table:summary-discard}).
\textit{Panel C:} satellites outside the former structure. 
In these panels, $\vec{J}_{\rm orb}$ are considered as axial vectors  
and are represented in just one hemisphere.
In addition, orbital poles have been rotated $60$ degrees in longitude 
with respect to Panel A for a clearer visualization.
Satellite IDs are specified  using  the same symbols and colors  as in Panel A.
\textit{Panel D:} Time evolution of the central galaxy's spin vector $\vec{J}_{\rm disk}$ relative to a fixed reference frame, both for  Aq-C$^\alpha$ (triangles) and PDEVA--5004 (circles). Different colors stand for different output times, as encoded in the  bar.
}
\label{fig:Aitoff_AqC_ALL}
\end{figure*}

\subsection{Satellite identification}
\label{sec:satIdent}
In both simulations we identify satellites at two times, $z=0$ and $z\sim0.5$ (T$_{\rm uni}$ = 8.66 Gyr for Aq-C$^{\alpha}$ and 8.68 Gyr for PDEVA-5004, respectively). The latter time is used to include satellites that may end up accreted\footnote{Satellite `accretion' stands for the (total or partial) incorporation of the  baryonic satellite mass into the central disk, causing  the individual satellite to vanish.}
 by the disk and do not survive until $z=0$.
We define satellites as all objects  with stars  ($M_\star>0$) that are bound to the  host galaxy within any radial distance.
To prove that a given object is a satellite galaxy (i.e. a bound object), we have computed its orbit.

In the case of Aq-C$^{\alpha}$, 
structures and substructures were identified using a Friends-of-Friends algorithm and the 
SubFind halo finder  \citep{Springel01}. 
We used the particle IDs  to
 trace the time evolution of the selected satellites and build merger trees.
Satellites in PDEVA-5004 were selected as sets of bound particles using  \texttt{IRHYS}\footnote{Simulation visualization and analysis tool developed by H.\,Artal. Not publicly available.}, which  we then followed in time.

The total number of satellites is of 
34  (35) in Aq-C$^{\alpha}$ (PDEVA-5004).
 Of these,  32  (26)
survive until $z=0$
\footnote{We note that we have improved our satellite tracking method in  Aq-C$^{\alpha}$, resulting  in an extension of the survivability of a  few   satellites further  in time than what presented in Paper II. }.
Relevant satellites to this study will be addressed throughout the paper with an identification code (see for example in Table \ref{table:summary-discard}).
Satellites in Aq-C$^{\alpha}$   (PDEVA-5004) show baryonic masses ranging between 
$M_{\rm bar} = 8.5 \times 10^6 - 8.9 \times 10^8\,\rm M_\odot$
($M_{\rm bar} = 3.9 \times 10^7 - 1.8 \times 10^8 \,\rm M_\odot$).
A discussion on the satellite mass distributions is given in Paper II, where it is shown that mass is not a  property
biasing satellite membership to thin planar distributions \citep[see also][]{Collins15}.

\subsection{Orbital properties}\label{sec:orbprop}
Satellites in both simulations present a diversity of orbital histories.
The majority have regular orbits whose apocentric and pericentric distances do not change appreciably during the analyzed time intervals. Some of them show long periods and large apocentric distances, larger than the virial radii in some cases (i.e., backsplash satellites). 
Finally, a small fraction of satellites 
see their  apocenters gradually reduced as they
end up accreted by the central galaxy's disk.
Regarding satellites with late incorporations to the system,
we find that 
several Aq-C$^\alpha$ satellites show first pericenters later than T$_{\rm uni}=10$ Gyr 
(including the MD)
 while all of PDEVA-5004's satellites are fully incorporated to the system by that time.

As mentioned in Paper II, a relevant fraction of satellites 
show pericentric coincidences at some particular cosmic times. In the short periods around these coincidences, an accumulation of satellites occurs at short distances from the main disk galaxy center, enhancing the strength of tidal effects and other possibly disturbing phenomena. 
Such orbital events  may have  consequences in the evolution of the individual orbital angular momentum of satellites as we will note throughout the paper.
In PDEVA-5004 particularly,  several satellites end up accreted by the disk in a short time interval around $T_{\rm uni}\approx 10$ Gyr, after one such pericenter coincidence.

\subsubsection{Evolution of the orbital angular momentum  with cosmic time}\label{sec:jorbevo}

We have computed the  orbital angular momentum vector $\vec{J}_{\rm orb}$ of each satellite at each timestep.
For a given satellite, this is the orbital angular momentum of
its center of mass 
relative to the host disk galaxy's center of mass.
Figure~\ref{fig:jsat_AqC}   
shows the time evolution of the directions of $\vec{J}_{\rm orb}$ vectors 
(i.e., orbital poles) for
the complete  satellite set identified in the Aq-C$^\alpha$  simulation.
Specifically, we show the evolution starting from high redshift of  $\vec{J}_{\rm orb}$'s  angular components $\theta,\phi$
 relative to axes that are kept fixed along cosmic evolution.
In the left panels we  see that, for many satellites,  the orbital poles 
are  roughly conserved 
along the slow phase of mass assembly, 
 while this conservation is not that clear for other satellites, see right panels in this Figure. 
Noticeably, T$_{\rm vir}$ is not a particular time concerning orbital pole evolution, as no discontinuity or change of behaviour in the orbital pole directions shows up at that moment. 
 Indeed, some satellites  roughly maintain their pole directions
from T$_{\rm uni} \sim 4$ Gyr onwards.
At T$_{\rm uni}$ lower than $\sim$ 2 Gyr, most satellite pole directions become noisy, as expected.

It is worth noting that the massive dwarf capture happening at T$_{\rm uni} \simeq$ 11.5 Gyr does not have a relevant impact on the $\vec{J}_{\rm orb}$ behaviour of the original satellites of the system, and, consequently, on the kinematic structures they form (see next sections).

To better illustrate and quantify the information on orbital pole changes,  we use Aitoff projections.
Figure~\ref{fig:Aitoff_AqC_ALL}A  shows the poles of the entire 
Aq-C$^\alpha$ satellite sample plotted from T$_{\rm vir}$ onwards. 
Satellite identities are coded by colors and  each point stands for a simulation output, i.e., different times. 
In Figure~\ref{fig:Aitoff_AqC_ALL}A we see that while some satellites have their representative points
corresponding to different timesteps
 very close to each other (e.g. satellites \#341, 29, 347), 
other span a large angular distance on the sphere (e.g. satellite \#360). The first are examples of satellites whose orbital poles  are \textit{conserved}, while the  last  do not conserve orbital angular momentum.

In the quest for persistent, kinematically-aligned satellite systems, the second necessary  ingredient is the \textit{clustering} of a high fraction of  conserved poles, belonging to different satellites. 
Examples of this behaviour can also be found in Figure~\ref{fig:Aitoff_AqC_ALL}A. We will take profit of this behaviour to determine, in Section \ref{sec:methodsims},
the axes of maximum satellite co-orbitation.

\subsubsection{Satellite orbital plane orientation with respect to the central disk galaxy}\label{sec:satorientation}
Figure~\ref{fig:Aitoff_AqC_ALL}D
shows, for each simulation, the time-evolution of the orientation of the central disk galaxy's spin vector, $\vec{J}_{\rm disk}$, relative to a fixed reference frame.
In the specific case of Aq-C$^\alpha$, in this occasion we use a reference frame that is oriented such that 
 at high redshift ($z\sim2.5$; $T_{\rm uni}\sim2.5$ Gyr)  the XY plane (located at latitude $\theta=0^\circ$ in the diagram)
 contains the proto-disk of the central main galaxy.

It is evident that each   $\vec{J}_{\rm disk}$ changes its direction  as the simulations evolve. 
In particular, Aq-C$^\alpha$'s disk axis flips from normal to the XY plane at high $z$  to roughly  perpendicular to its former direction, with  $\theta$ smoothly changing $\Delta\theta\sim90^\circ$ in all. 
After that, the host disk  remains 
normal  to the formerly mentioned  XY plane, 
but rotating very slightly around an axis roughly normal to this  plane.
In turn, PDEVA-5004's disk spin vector 
moves very little 
since T$_{\rm vir}$
and only up to $T_{\rm uni}\approx10.5$ Gyr.
 
 We have studied the angle 
 formed by $\vec{J}_{\rm orb}$ and $\vec{J}_{\rm disk}$ at each timestep for all satellites.
In both simulations, we have satellites spanning a wide range of   $\alpha(\vec{J}_{\rm disk},\vec{J}_{\rm orb})$ values. 
Angle variations with time are rather smooth in general. Some satellites show long-range  patterns in the evolution of  $\alpha(\vec{J}_{\rm disk},\vec{J}_{\rm orb})$ that are mainly due to the motions of the disk relative to the  fixed reference frame.  
On the other hand,
most satellites that will end up accreted by the disk show a gradual decrease of  $\alpha(\vec{J}_{\rm disk},\vec{J}_{\rm orb})$,
such that their accretion happens in the plane of the disk. 
Finally,
in both simulations there is an important fraction   of satellites with regular, periodic orbital behaviour that are on near-to-polar orbits, this is, with  $\alpha(\vec{J}_{\rm disk},\vec{J}_{\rm orb})\sim90^\circ$. 
We will study the relative orientation of satellites
forming kinematic planar structures
 in more detail in Sections~\ref{sec:orientation} and \ref{sec:PropKineSat}.

%
%

\section{Identifying Kinematically-Persistent Planes of satellites: method}
\label{sec:results}

Identifying kinematically-persistent planes involves two steps: i), determining  the axes around which the fraction of co-orbiting satellites is a local maximum at each moment within a given time interval, and,
 ii), identifying the specific, fixed set of satellites that persistently co-orbit these axes along this time interval. We describe our method in the following subsections.

\subsection{Identifying directions of maximum satellite co-orbitation}
\label{sec:methodsims}
Scanning over the projections of $\vec{J}_{\rm orb}$ directions traced on the sphere in search for orbital pole overdensities, at each timestep, appears as 
a direct, straightforward method  to look for the direction around which most orbital poles are enclosed within an aperture. However, $N_{\rm tot}$ is generally not high enough for this direct method to converge and produce meaningful results.

As we aim at determining   sets of satellites showing orbital pole  clustering that is stable along a long interval of time,
a second justifiable method consists  in considering  the  satellite  poles  
corresponding to this time interval altogether.
This is, for each satellite, its orbital pole directions corresponding to all the timesteps within the period under consideration, are projected on the sphere.
Then, we look for 
 clustering in this sample of directions
by scanning
  the sphere  just once  with a given aperture angle $\Delta_{\rm scan}$
  in search for an axis around which poles accumulate. 

This protocol will be hereafter termed  the  \textit{`Scanning of Stacked  Orbital Poles Method'}, the reason being that, for each satellite, the values of its corresponding orbital poles at each simulation output or timestep, have been `stacked'. By using a stacked sample of orbital poles, the number of  $\vec{J}_{\rm orb}$ directions is multiplied by the number of timesteps within the period considered, and convergence
in finding clustering
 is ensured for both simulations.

Note that the longer the periods of time considered, the higher the number of $\vec{J}_{\rm orb}$ directions to combine together.  
Ideally, the `stacking' time-intervals should be  as long as possible.
We use as default stacking interval from T$_{\rm vir}$ to T$_{\rm z=0}$\footnote{We have checked that the maxima in the orbital pole  clustering our method returns is robust under extensions of the stacking  interval down to T$_{\rm uni}\sim 4$  Gyr.}.
 A limitation 
to this long period, however, would be 
  the  appearance of a dynamical configuration in the system that could possibly spoil or disolve the stable clustering of a subset of orbital poles.
Since we observe such potentially disturbing dynamical events in both simulations we analyze here, we have considered splitting the default T$_{\rm vir}$ - T$_{\rm z=0}$ range into two sub-intervals.

In the case of the Aq-C$^\alpha$ simulation, the system is free from suffering  potentially disturbing dynamical events until T$_{\rm uni} \sim $ = 11.5  Gyr  ($z=0.18$).  Soon after, the system enters into an interaction phase,
due to the approach and capture of the MD system.
In consequence, the T$_{\rm vir}$ - T$_{\rm z=0}$ range has been split in two sub-intervals separated by T$_{\rm uni}$ = 11.5  Gyr. Results obtained using these two time intervals  have been compared with those obtained 
when using the full T$_{\rm vir}$ - T$_{z=0}$ range. 
As expected from the behaviour of satellite orbital poles   around $T_{\rm uni} \sim 11.5$  Gyr (i.e., no major impact from the capture, see Figure~\ref{fig:jsat_AqC}), results  for 
the axes of maximum co-orbitation obtained in either case are essentially the same.
Therefore in what follows we will focus on our results from the analysis of Aq-C$^\alpha$ 
from T$_{\rm vir}$ to T$_{\rm z=0}$.
On the other hand,
given the particular dynamical situation the   PDEVA-5004  system suffers around 10 Gyr
(when several satellites are accreted by the main galaxy),
we have considered two  stacking sub-intervals in this case.

To improve  the method, some details are in order.   
Satellites 
whose orbital pole conservation is not good enough 
contribute noise (see Section~\ref{sec:jorbevo} and Figure~\ref{fig:Aitoff_AqC_ALL}A). 
In order to minimise noise, 
satellites that end up accreted by the central galaxy, as well as
 satellites whose orbital poles sweep an aperture angle higher than 40$^\circ$ in the total period analyzed,
 have been discarded in the stacking procedure. 
Additionally, in the specific case of Aq-C$^\alpha$, the 
late-captured
 massive dwarf and its satellites  are not included in the stacking procedure, because their  membership to the system cannot be guaranteed  until T$_{\rm uni} \sim$ 12 Gyr.

To still increase the local number density of directions and make it easier to identify axes of co-orbitation, in the stacking procedure  the satellite orbital poles  $\vec{J}_{\rm orb}$ have been considered as \textit{axial} vectors, i.e., their sense of rotation has not been taken into account.
In this case, each of them intercepts twice the unit sphere at diametrically opposite positions, 
that are indistinguishable for stacking purposes.
Hence, the information given in Figure~\ref{fig:Aitoff_AqC_ALL}A can be communicated using a unique hemisphere. 
As an illustration of our methodology,
in Panels B \& C of Figure~\ref{fig:Aitoff_AqC_ALL} the same information is given  using a unique hemisphere, and, additionally, tracks belonging to satellites contributing noise to the stacking procedure have been removed. 
 
The resulting axis of maximum satellite co-orbitation,
$\vec{J}_{\rm stack}$, identified for the Aq-C$^\alpha$ satellite system, is shown as magenta `x' crosses in Figure~\ref{fig:Aitoff_AqC_ALL}.
The precise location of the $\vec{J}_{\rm stack}$ axis depends on the scanning aperture $\Delta_{\rm scan}$ used.
In Figure~\ref{fig:Aitoff_AqC_ALL}  we plot  results for 
our choice of
$\Delta_{\rm scan} =  34^\circ$. 
Results using a slightly larger value of $\Delta_{\rm scan} = \alpha_{\rm co-orbit} =  36.87^\circ$ have also been analyzed, and are very similar (see next Section~\ref{sec:MemberDet} and Table~\ref{table:summary-discard}). 
Lower $\Delta_{\rm scan}$ values have also been explored, but 
the resulting  $\vec{J}_{\rm stack}$ axes
 lead to collimated orbital pole groups consisting of a low number of satellites, making them  uninteresting for our purposes here.
A similar analysis applied to the PDEVA-5004 system returns two $\vec{J}_{\rm stack}$ axes, see
Figure~\ref{fig:PDEVA-Jsatck} in Appendix~\ref{appendix1}.

Figure~\ref{fig:Aitoff_AqC_ALL}(B \& C)  show that the  $\vec{J}_{\rm stack}$ axis found indeed points in a direction where the poles of many Aq-C$^\alpha$ satellites cluster.  
To emphasize these results, Panel B shows the poles of the KPP satellite members (identified in next Section~\ref{sec:MemberDet}), while those satellites not contributing to this structure are drawn in Panel C.

In Paper I an axis of maximum satellite co-orbitation  in the Milky Way was  determined  from  kinematic data of satellites at $z=0$. As no timestep stacking is possible in that case, a trick was used that  increases  the number of directions projected on the sphere, keeping track of the satellite orbital poles' space organization. 
The so-called \textit{`3$\vec{J}_{\rm orb}$-barycenter Method'} and its extension to determine KPPs in simulations  is described here in Appendix~\ref{appendix2}.

\begin{table*}
\centering
\scriptsize
\caption{Summary Table for groups of kinematically-coherent satellites identified in the simulations by means of the \textit{Scanning of Stacked Orbital Poles method} (see Section~\ref{sec:results}).
Columns give the satellite IDs ordered according to $N_{\rm co-orb}$ (see text).
The scanning apertures $\Delta_{\rm scan}$ used to determine the $\vec{J}_{\rm stack}$ axes are given in the Table headers, where it is also indicated whether one ($1T$) or two ($2T$)  stacking time intervals have been considered.   
IDs colored red stand for satellites that rotate in one same sense within the KPP. Those in black rotate in the contrary sense relative to the former.
Numbers in parenthesis  placed at the $i$-th row  give the ratio of satellites rotating in one sense over those rotating in the opposite, for satellites  whose IDs are in the first $i$ rows.
}
\begin{tabular}{| l l | l l | l  |}
\toprule
\midrule
        \multicolumn{2}{c}{Aq-C$^\alpha$}    & \multicolumn{2}{c}{PDEVA-5004 (1)}   & \multicolumn{1}{c}{PDEVA-5004 (2)}    \\

\midrule
\midrule
         $\Delta_{\rm scan}=36.87^\circ$  &  $\Delta_{\rm scan}=34^\circ$  &  $\Delta_{\rm scan}=34^\circ {\rm or\,} 36.87^\circ, 1T$  & $\Delta_{\rm scan}=34^\circ {\rm or\,} 36.87^\circ, 2T$ &  $\Delta_{\rm scan}=34^\circ {\rm or\,} 36.87^\circ, 1$ or $2 T$     \\
\midrule
\midrule
    506       &   506               & 5          & 40        &  9            \\
  \red{346}    &   \red{346}          & \red{10}   & \red{10}  &\red{13}       \\
  \red{14}     &  \red{14}               & 11         & 11        &  19           \\
  \red{341}     &   \red{341}                & 26         & 26        &\red{23}       \\
      473       & 478          & \red{27}   & \red{27}  &\red{20}       \\
  \red{353}       &  \red{353}               & \red{29}   & \red{29}  &\red{16}       \\
    474          &  \red{352}          & 40         & \red{28}  &  1 (4/3)      \\
    29           &  29        & \red{28}   & \red{8}   &               \\
    345          &   345               & \red{8}    & 5         &               \\
 \red{21}        &  349           & 18  (5/5)  &\red{22} (6/4) &           \\
   349   (5/6)     &\red{21} (6/5) &         &                 &               \\
   478  (5/7)     &   \red{38} (7/5)    &         &                 &               \\
    \red{20}   (6/7)  &   474 (7/6)   &         &                 &               \\
\midrule
\bottomrule
\end{tabular}
\label{table:summary-discard}
\end{table*}

\subsection{KPP satellite membership determination}
\label{sec:MemberDet}
Results in the previous subsection indicate that satellite orbital poles cluster in preferred directions.
Here we use these co-orbitation axes to single out satellites that define KPPs.

To search for and define  satellite groups  associated to a given  $\vec{J}_{\rm stack}$ axis, we order satellites by decreasing  number of timesteps  within the analyzed period of time, $N_{\rm co-orb}$, that they cluster around their corresponding $\vec{J}_{\rm stack}$ axis.
This is, $N_{\rm co-orb}$ represents the number of timesteps in which  a satellite's poles fall within an angle of $\alpha_{\rm co-orbit}$ from the  $\vec{J}_{\rm stack}$ axis, marked with a circle in Figures~\ref{fig:Aitoff_AqC_ALL} and \ref{fig:PDEVA-Jsatck}. 
To make use of all the available information about pole conservation (see Figure~\ref{fig:jsat_AqC}) we consider the time interval from T$_{\rm uni}=4$ Gyrs to T$_{\rm z=0}$.  
The number of satellite members in KPPs has been determined by the criteria that their poles are within the co-orbiting circle for at least a 50\% of their outputs\footnote{This is equivalent to requiring poles are within the co-orbiting circle for at least half of the time interval we consider.}. This gives
 13 satellites in the Aq-C$^\alpha$ KPP, and 
 10 (7) satellite members for KPP1 (KPP2) in PDEVA-5004.
In Section~\ref{sec:results} we explore whether or not the  properties of KPP planes as positional configurations are robust when a more restrictive condition (i.e., a lower number of satellites in KPPs)  is imposed.

Table~\ref{table:summary-discard} gives the KPP members for Aq-C$^\alpha$. The first (second) column shows results where a $\Delta_{\rm scan}=36.87^\circ$ ($34^\circ$)  has been used to determine the stacking axes.
Satellite membership differs only very slightly (i.e., just a couple of satellites) between columns first and second of  Table \ref{table:summary-discard}.

Satellites whose  IDs are colored red in Table~\ref{table:summary-discard} rotate in one same sense within the KPP. Those in black rotate in the contrary sense relative to the former.
Numbers in brackets at  the $i$-th row,  give the ratio of satellites, among those   placed at the first $i$ rows, rotating in a sense over those rotating in the contrary.
These ratios change  from row to row (i.e., they depend on the number of satellites we take in a given KPP), 
fluctuating around a 50\%. 
However, they further change when including the MD system at $z=0$, see Section~\ref{sec:late-capture}.

 We proceed in the same way to identify KPP members among PDEVA-5004 satellites.
Results are given in the third, fourth and fifth columns of Table \ref{table:summary-discard}. KPP  member satellites  have been searched for either using a unique stacking interval
(results marked $1T$),
or two of them, breaking at  T$_{\rm uni}$ = 10.5 Gyr 
(marked $2T$).
The satellite group corresponding to the main  $\vec{J}_{\rm stack}$ axis (KPP1)
differs in just 1 satellite member between the $1T$ or $2T$ cases.
As for the second group (KPP2), it  turns out that no differences exist either in the satellite   membership or ordering, when one or two stacking intervals are used. Results are given in the fifth column for both cases. 
Table \ref{table:summary-discard} shows results corresponding to $\Delta_{\rm scan} = 34^\circ$ or 
$\Delta_{\rm scan} = 36.87^\circ$, as KPP1 and KPP2 satellite members are exactly the same in both cases.

Results obtained by using the `3$\vec{J}_{\rm orb}$-barycenter'  method are given in Table~\ref{table:summary-3JB}.
No differences in KPP membership have been found when using one method or the other  in the PDEVA-5004 system, and just 1 satellite varies in the case of Aq-C$^\alpha$.
This satisfactory consistency of results using both methods when applied on the two simulations is reassuring and demonstrates our findings are robust.

Each of these groups of kinematically-coherent satellites
 are candidates to form good-quality \textit{positionally-defined} planes. 
In Section~\ref{sec:KineSat}   we study the
time evolution of their spatial characteristics.

\subsection{Quantification of the degree of satellite co-orbitation around the axes identified }
\begin{figure*}
\centering
\includegraphics[width=0.495\linewidth]{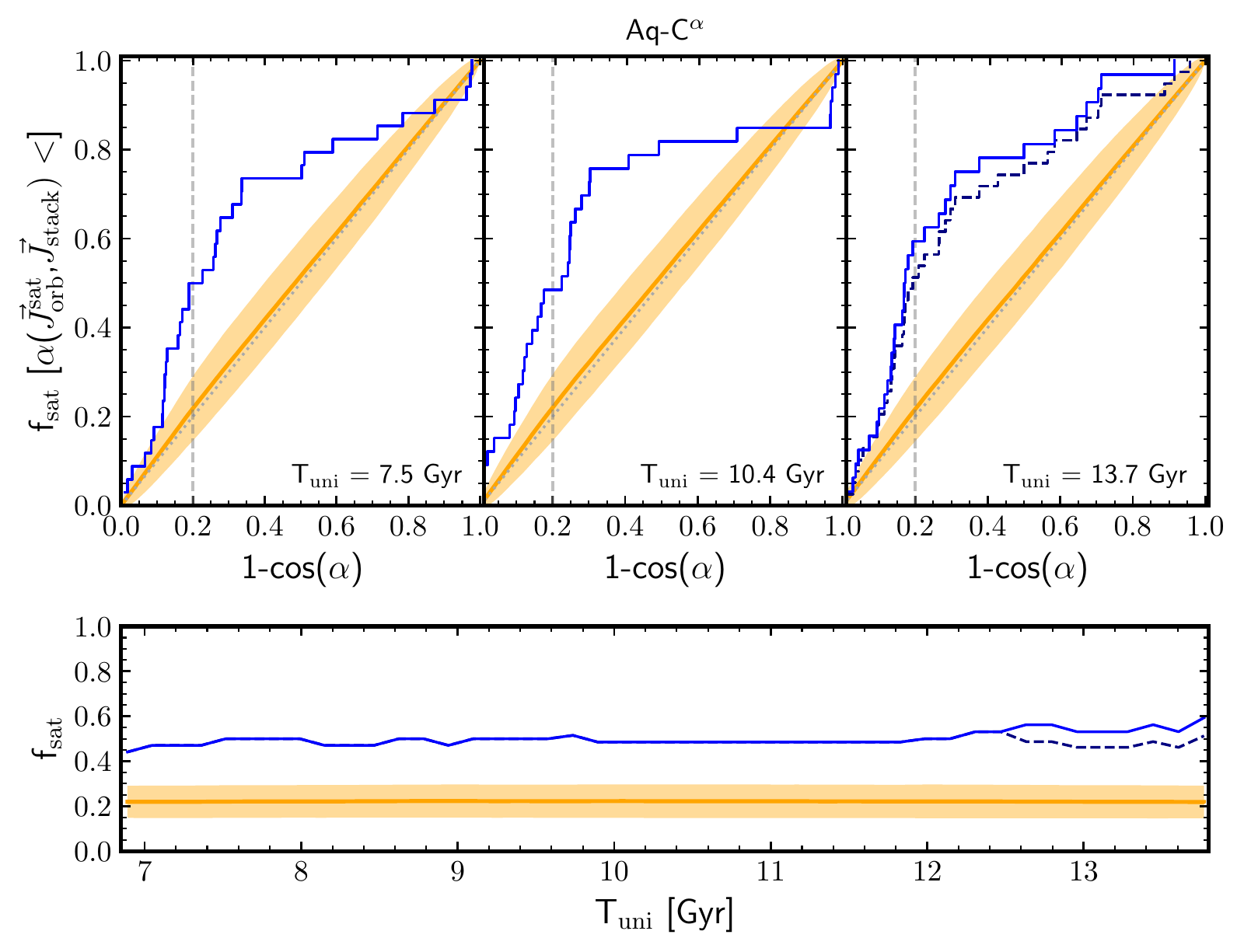}
\includegraphics[width=0.495\linewidth]{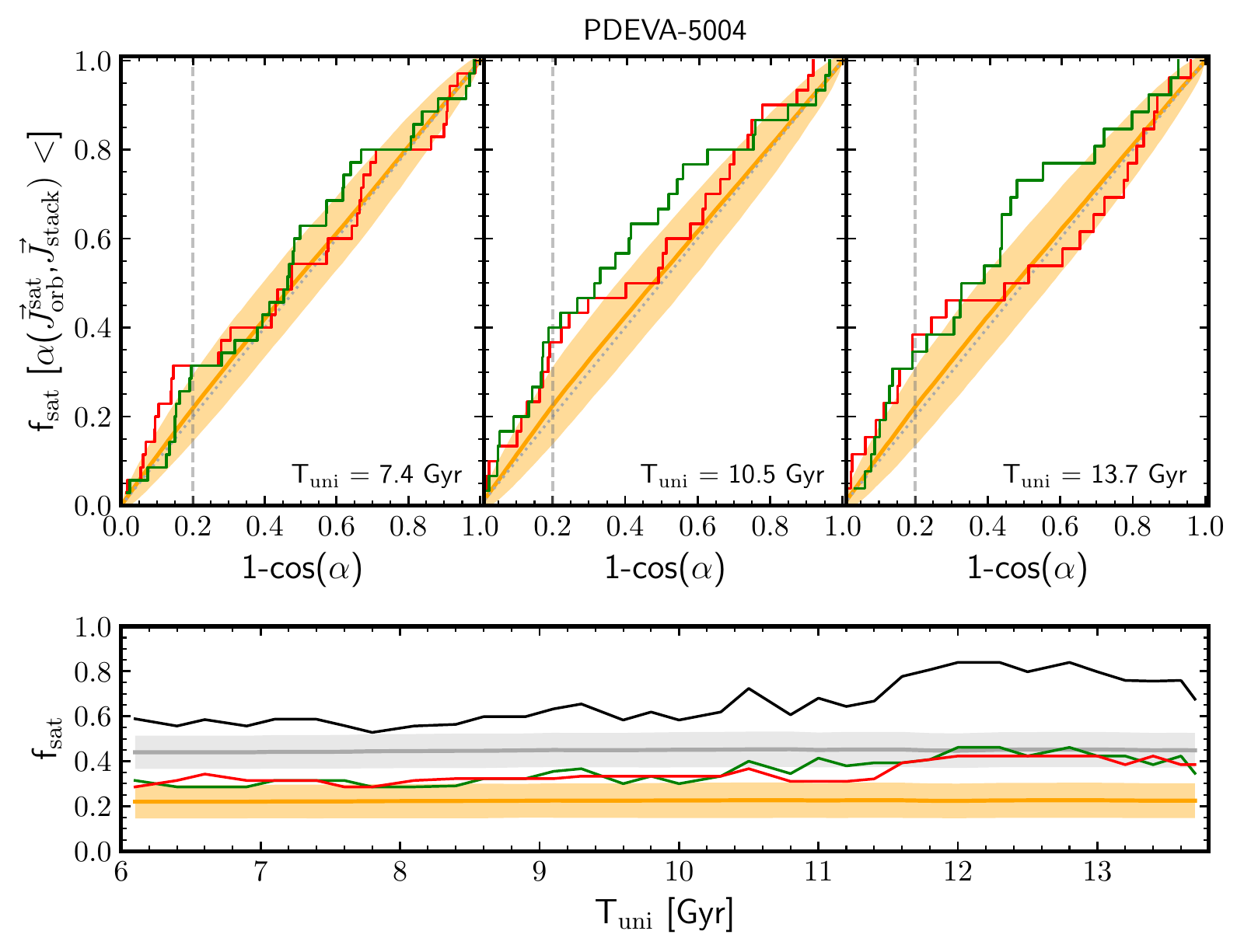}
\caption{\textit{Upper panels}:  Fraction of satellites with orbital poles  $\vec{J}_{\rm orb}$ enclosed within an angle $\alpha$ 
measured from the co-orbitation axes $\vec{J}_{\rm stack}$, at given timesteps.
Each panel indicates the Universe age $T_{\rm uni}$.
Left: Aq-C$^\alpha$; right: PDEVA-5004 (measured from main $\vec{J}_{\rm stack}$ axis in red, from second axis in green).
The orange lines  with   shade  band give the results  when the method is applied to isotropized configurations of orbital poles (see Appendix~\ref{appendix3}), specifically, the averages and 1$\sigma$ dispersion band  over N$_{\rm random}$ = 5000  randomized realizations.
Gray dotted diagonal lines stand for  the analytical result for an homogeneously sampled sphere.
A  vertical line marks the satellite co-orbitation criteria adopted in this work, i.e., an angle of $\alpha_{\rm co-orb}=36.87^\circ$ (see text). 
The dashed line for Aq-C$^\alpha$ represents the situation when including the lately-captured massive dwarf and its sub-satellites.
\textit{Lower panels}: Time evolution of the fraction of satellites with orbital poles within an angle of $\alpha_{\rm co-orb}$ from the respective $\vec{J}_{\rm stack}$ axes (i.e. the fraction of `co-orbiting' satellites).
For PDEVA-5004, a black line shows the total co-orbiting fraction by summing up contributions from the red and green lines, and a gray line and band indicate the corresponding isotropization mean and scatter to compare it to. 
}
\label{fig:unocos}
\end{figure*}

Figure~\ref{fig:unocos} helps us 
gain a deeper insight into the co-orbitation behaviour of satellites.
In the upper panels we  show, for three timesteps,  
the fraction of satellites (relative to the whole sample) with orbital poles within a certain angle $\alpha$ measured from the $\vec{J}_{\rm stack}$ axis: this is, the Cumulative Distribution Function (CDF) for the satellite pole angular distances relative to $\vec{J}_{\rm stack}$.
In the bottom panels we give the evolution,
after the halo collapses at T$_{\rm vir}$,
 of the fraction of co-orbiting (within $\alpha_{\rm co-orbit}=36.87^\circ$) satellites as measured from $\vec{J}_{\rm stack}$.
Results for Aq-C$^\alpha$ (PDEVA-5004) are shown in the left (right) panels, see legend for details.

A measure of the bias the method gives rise to
can be obtained by comparing
 the difference between orange and gray dotted diagonal lines in the top panels of Figure~\ref{fig:unocos}.
We see that it is unimportant.

Figure~\ref{fig:unocos} shows that Aq-C$^\alpha$ and PDEVA-5004 present different degrees of orbital pole collimation.
In first place, we note that satellite orbital poles in the two simulations are never randomly distributed at small angles (i.e., $<\alpha_{\rm co-orbit}$; $1-\cos(\alpha)<0.2$): at any time there is a relevant fraction of ``co-orbiting" satellites out of the total.
Aq-C$^\alpha$ shows a remarkably high fraction of co-orbiting satellites at all timesteps. CDFs  exceed the signal for isotropized distributions at $1-5\sigma$ across different angles. In detail, in the few timesteps illustrated,  the orbital pole collimation is higher at  angular scales $\alpha>26^\circ$ ($1-\cos(\alpha)>0.1$) than at smaller scales.

On the other hand, PDEVA-5004's 
small scale degree of collimation is similar  for both  $\vec{J}_{\rm stack}$ axes, of around $>1\sigma$ for T$_{\rm uni}<9$ Gyr, and of about $>2-3\sigma$ at later times. The large scale collimation ($1-\cos(\alpha)>0.5$) is however low, showing a distribution compatible with isotropy at some timesteps.

We condense the previous results on time behaviour in the bottom panels of Figure~\ref{fig:unocos}.
The fraction of co-orbiting satellites is mostly maintained with only small fluctuations for Aq-C$^\alpha$ 
with an average of
$\sim0.50$ (or $\sim0.49$ when including the MD system at low $z$), and reaching higher values at given moments:  i.e., $\sim$half of Aq-C$^\alpha$'s satellite population is co-orbiting on a common orbital plane during the complete period studied. 
The fraction for both $J_{\rm stack}$ axes in PDEVA-5004 increases
as cosmic time evolves.
The main (second) co-orbitation axis in PDEVA-5004 defines a direction with  an average co-orbiting fraction of $ \sim 0.35$ ($\sim 0.36$):
this is, in total
$ \sim 70\%$ of PDEVA-5004's satellites show a high kinematical organization considering both planes, the fraction reaching up to a $80\%$ of them at particular times (see black line in bottom panel).
Comparing with the orange line and shaded band in the bottom panels, we see that the clustering of orbital poles at the  $\alpha_{\rm co-orbit}$ scale around each $\vec{J}_{\rm stack}$ direction exceeds that expected from an isotropized distribution. The excess is less important ($\sim1\sigma$) for PDEVA-5004  at T$_{\rm uni}\lesssim 9$ Gyr, and very relevant for Aq-C$^\alpha$ (some $5\sigma$ along the period analyzed).\footnote{One may examine how these simulated orbital pole distributions and co-orbiting fractions
compare to those measured for the Milky Way satellite sample
at $z=0$. We refer the reader to figure 13 in Paper I. A comparison between simulations and the MW
satellite sample shows results are similar. 
We note, however, that the MW is different from the simulated galaxies analyzed here in
different respects. Therefore no quantitative agreement should be expected,
but just some qualitative similarity as the possible consequence
of the same physics.}


\section{Planes of kinematically-coherent satellites}
\label{sec:KineSat}

\begin{figure*}
\centering
\includegraphics[width=0.32\linewidth]{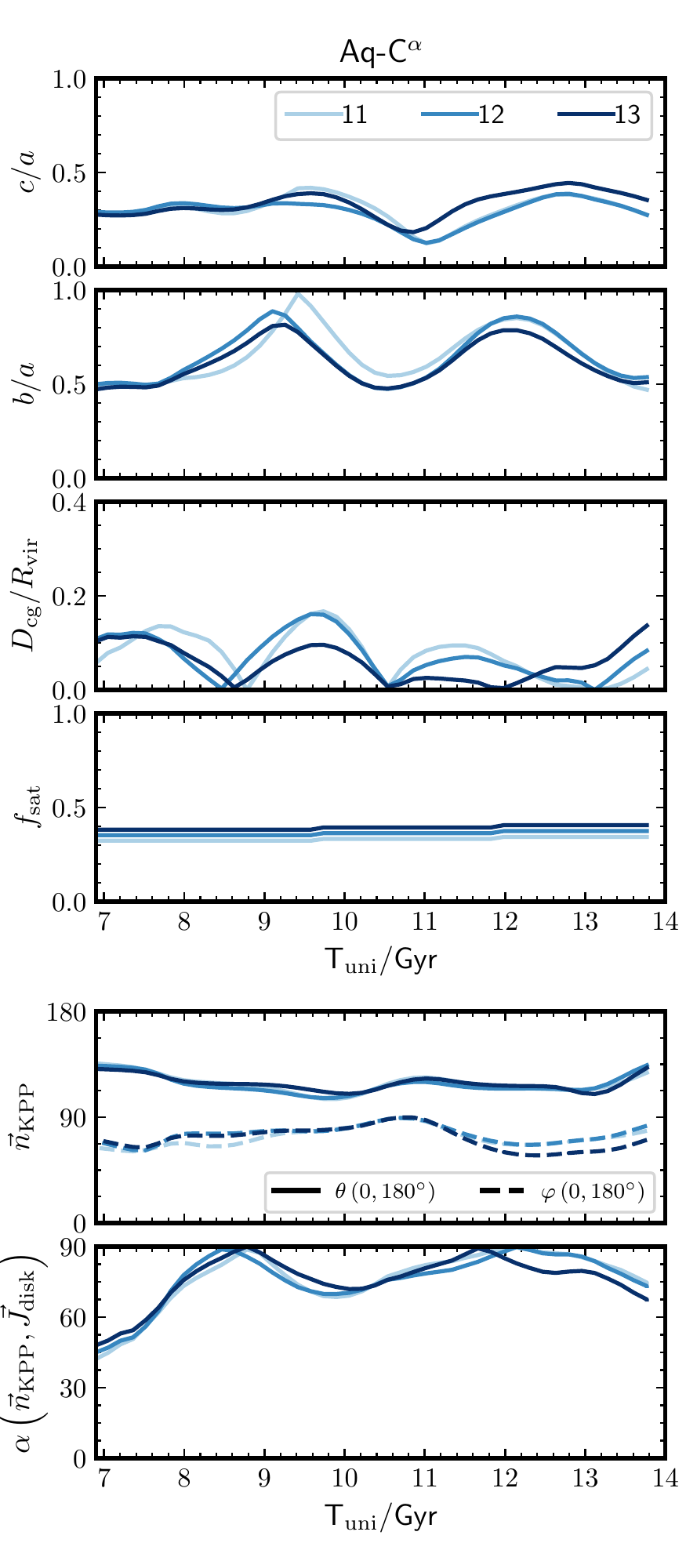}
\includegraphics[width=0.32\linewidth]{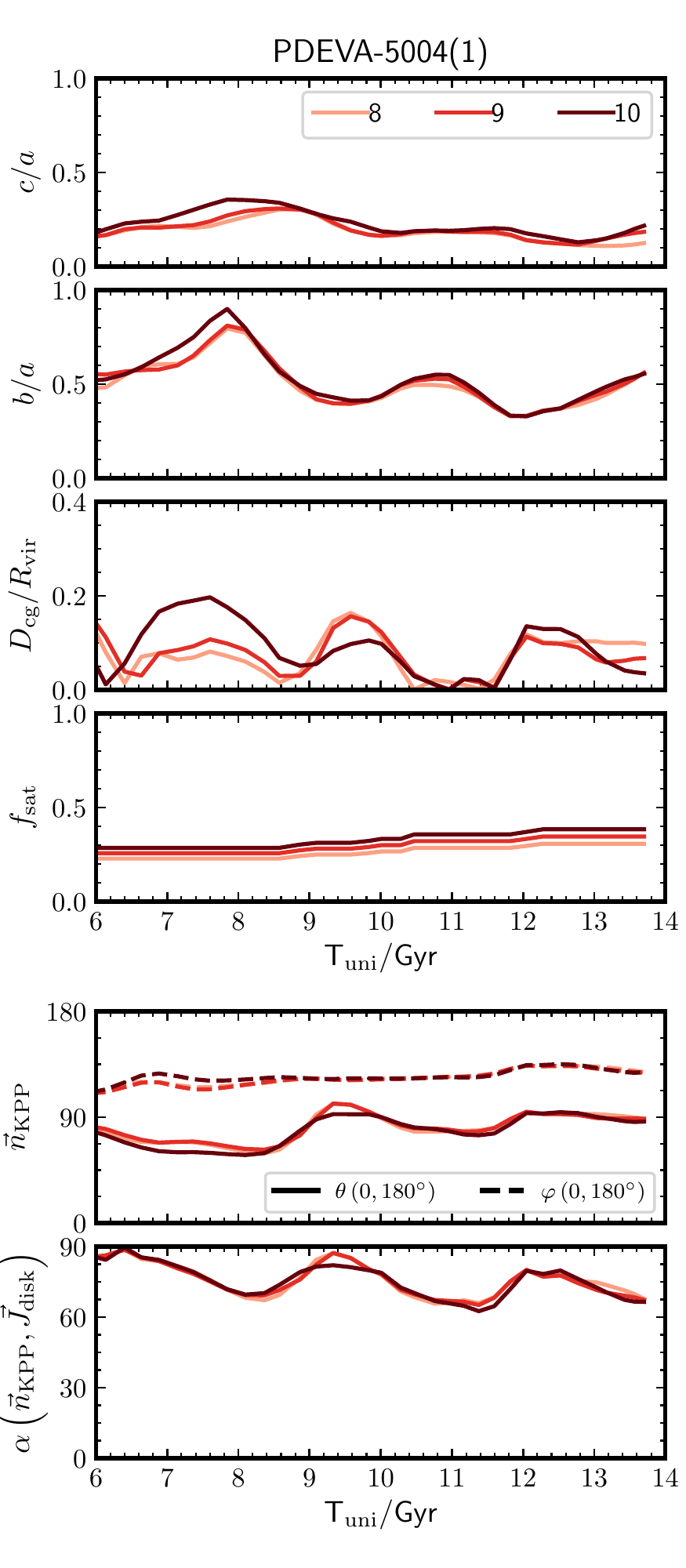}
\includegraphics[width=0.32\linewidth]{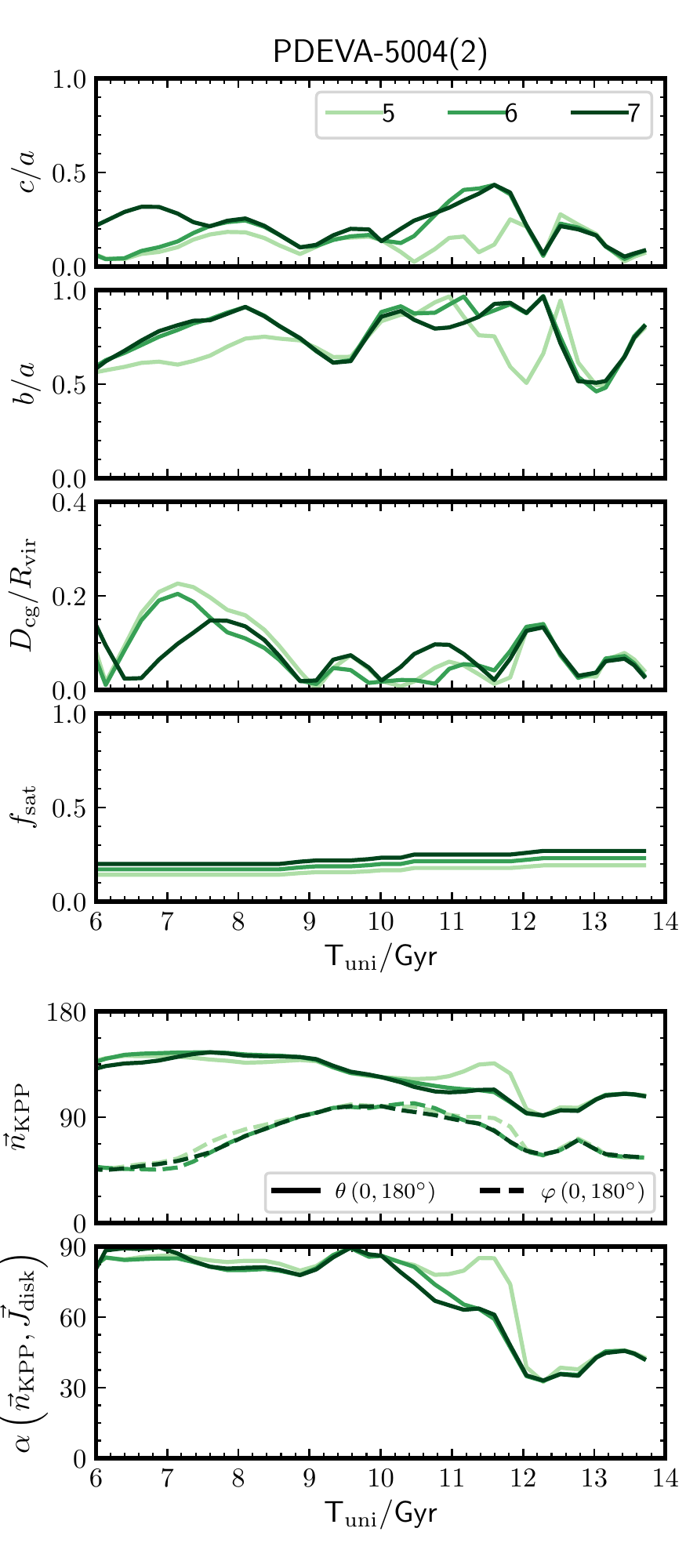}
\caption{Time evolution of the planes formed by the groups of kinematically-coherent satellites identified in each simulation, see Table \ref{table:summary-discard}. Left: Aq-C$^\alpha$. Middle: PDEVA-5004 (1) (group corresponding to the main $\vec{J}_{\rm stack}$ axis). Right: PDEVA-5004 (2)  (group corresponding to the second axis).
Panels in the upper blocks show the ToI plane parameters (See Section~\ref{sec:quality}). 
Panels in the lower blocks show 
the time evolution of the normal vectors to the KPPs, $\vec{n}_{\rm KPP}$ in spherical $\theta,\varphi$ angular components with respect to a fixed reference frame;
and
 the angle $\alpha$ formed between $\vec{n}_{\rm KPP}$  and the spin vector of the central galaxy $\vec{J}_{\rm disk}$.
Colored lines of different shades stand for subsamples of satellites  out of the total kinematically-coherent group with different number of satellites $N_{\rm sat}$, as indicated in the legend, 
following the $N_{\rm co-orb}$ ordering from Table~\ref{table:summary-discard}.
}
\label{fig:kineplane-ori}
\end{figure*}


\subsection{Qualities of the KPPs}\label{sec:quality}
In order to analyze the \textit{quality} of sets of kinematic satellites  
as planar configurations in position space, we use the standard Tensor of Inertia (ToI) method \citep{Cramer}. Following this approach, an ellipsoid is fitted to the point positions of satellites and a planar configuration results, which can be described by:
\begin{itemize}[leftmargin=*,noitemsep,nolistsep,nosep]
\item $N_{\rm sat}$: the number of satellites in the subset (or, the fraction of the total number of satellites it represents $f_{\rm sat} \equiv N_{\rm sat}/N_{\rm tot}$).
\item $c/a$: the ellipsoid short-to-long axis ratio.
\item $b/a$:  the ellipsoid intermediate-to-long axis ratio.
\item $D_{\rm cg}/R_{\rm vir}$: the distance from the center-of-mass of the central galaxy to the plane, relative to the virial radius (a characteristic total size of the system).
\item $\vec{n}$: the direction of the normal vector to the planar configuration.
\end{itemize}

We consider as high-quality planes those that are thin,
 rather oblate 
 and populated (i.e., consisting of a relatively high fraction $f_{\rm sat}$ of the total amount of satellites existing at a given time), see Paper II for details.

The ToI analysis has been performed over the whole  satellite set listed in Table~\ref{table:summary-discard}, and then it has been repeated discarding the two last satellites listed there.
This is to test whether the 50\% $N_{\rm co-orb}$ criterion to determine the number of KPP satellite members results into sets whose properties in positional space are robust, or a more restrictive condition is needed.

Figure~\ref{fig:kineplane-ori} shows some relevant results 
 for the Aq-C$^\alpha$ (left column) and PDEVA-5004 (middle and right columns) simulations. 
 In particular, in the upper block of panels we plot the time evolution,
after halo virialization,
  of the $c/a$, $b/a$, $D_{\rm cg}/R_{\rm vir}$, and $f_{\rm sat}$ plane parameters.
In the lower block of panels, we plot the orientation of the planes.
Figure~\ref{fig:kineplane-ori} shows that all KPP plane characteristics vary with time
(see legends), but not largely.
In the case of Aq-C$^\alpha$, the plane remains thin ($c/a$ lower than $\sim0.3$) and oblate ($b/a>0.5$) for the bulk of the period analyzed. 
Near  $T_{\rm uni} \sim  10 $ Gyr there is a somewhat larger fluctuation of these parameters, with lower $b/a$ values, but these
increase later on.
On the other hand, $c/a$ and $b/a$ hardly vary between different $N_{\rm sat}$ groups,
confirming that the 50\% N$_{\rm co-orb}$ criterion used defines robust structures.

The distance to the central galaxy shows mostly low values   (below $\sim10\%$ of the virial radius),
however, it shows more fluctuations at different times.
We have confirmed that times when $D_{\rm cg}/R_{\rm vir}$ is relatively higher correspond to moments when the satellite distribution is heavily lopsided, with a higher fraction of satellites on one side of the central galaxy than on the other, dragging the fitting plane towards the former side.

Finally, the different $N_{\rm sat}$ groups of Aq-C$^\alpha$ kinematic satellites define planes that are highly populated relative to the total number of satellites of the system. In particular, they represent a  fraction of 
$0.32-0.41$ 
of the total.

PDEVA-5004's main plane of kinematically-coherent satellites (shown in red, middle panels) presents as well a remarkable thinness across time, with $c/a\sim0.2$, and decreasing as time passes. In general, the different quality indicators do not show any relevant fluctuations when $N_{\rm sat}$ changes,  an indication of a robust kinematic structure. 
More particularly, $b/a$ is not that close to unity, suggesting that the kinematically-persistent plane is not that oblate as in the case of  Aq-C$^\alpha$.
$D_{\rm cg}/R_{\rm vir}$ values fluctuate 
but show overall similar rather low values in both simulations.
The  kinematically-coherent satellites
represent a maximum fraction of up to $f_{\rm sat} \simeq 0.4$ of the total (for $N_{\rm sat}=10$), which occcurs at late times.

The second group of kinematically-coherent satellites identified in PDEVA-5004 gives rise as well to a planar structure in space (shown in green, right panels).
$c/a$ values show more fluctuations than for the main group  but are still very low during certain periods,
and $b/a$ is always very high, ratifying the oblateness of the structure.
Finally, this group represents a maximum fraction of $f_{\rm sat}\sim0.3$ with $N_{\rm sat}=7$.

In summary, from Figure~\ref{fig:kineplane-ori} it is clear how the subsets of kinematically-coherent satellites drawn from either simulation form  \textit{time-persistent planar} configurations, with a good quality during most of the analyzed periods.
Furthermore, these KPPs  attain a remarkably high quality at some specific timesteps.

We finish this subsection by noting that the positional-only analysis presented in Paper II  provided with higher quality planes of satellites in 
both simulations
at some redshifts than those presented here (i.e., more populated for equivalent $c/a$ values; see figure~9 and table~2 in Paper II). The extra satellites are usually interpreted as transient ones, as mentioned in Section~\ref{sec:intro}.
We compare positional- and kinematically-identified planes in Section~\ref{sec:KPPvsPosPla}.

\subsection{Orientation  of the `kinematically-persistent' planes of satellites}
\label{sec:orientation}
In this section we study the time evolution of the orientation of the KPPs, in relation to (a) a fixed reference frame, 
and (b) the plane of the disk of the central galaxy.

The lower block of panels in Figure~\ref{fig:kineplane-ori} show the evolution of the normal vector to the persistent plane, $\vec{n}_{\rm KPP}$, in spherical coordinates $\theta$ and $\varphi$ with respect to the fixed reference frame used for each simulation in Section~\ref{sec:jorbevo}.
An outstanding result is that, 
in both simulations, these orientations or angles do not change appreciably with variations of the number of satellites included in the plane, $N_{\rm sat}$. Moreover, 
the $\vec{n}_{\rm KPP}$ vectors remain approximately fixed in time, with angular components varying steadily and not by much.
 Indeed no important fast fluctuations of the direction are found for either simulation.

We also note  that all three $\vec{n}_{\rm KPP}$ directions across time are in rough consistency with their corresponding unique $\vec{J}_{\rm stack}$ directions.

The bottom panels of Figure~\ref{fig:kineplane-ori} show the angle formed by the $\vec{n}_{\rm KPP}$ normal vector and the  $\vec{J}_{\rm disk}$ spin vector of the central galaxy, $\alpha(\vec{n}_{\rm KPP},\vec{J}_{\rm disk})$.
Note that, as mentioned in Section~\ref{sec:satorientation} and shown in   Figure~\ref{fig:Aitoff_AqC_ALL}D, the central galaxies of both simulations  present their own motions within the
fixed reference frame.
Thus these results reflect the combined effect  of the motions of the kinematically-coherent satellites and of the disk.

In the case of Aq-C$^\alpha$, we find that $\alpha(\vec{n}_{\rm KPP},\vec{J}_{\rm disk})$ increases fast from virialization time until   $T_{\rm uni}\approx 9$ Gyr 
where it reaches $90^\circ$.
Beyond this time the central disk galaxy does not vary its orientation 
relative to the XY plane relevantly.
By $T_{\rm uni} \sim 11.5$ Gyr,  $\vec{n}_{\rm KPP}$ and $\vec{J}_{\rm disk}$ become perpendicular again
and in general remain roughly so until $z=0$, with small angle fluctuations  therefore only due to minor changes in the orientation of the $\vec{n}_{\rm KPP}$ vector.

PDEVA-5004's main persistent plane (red, middle column in Figure~\ref{fig:kineplane-ori}) is also roughly perpendicular to its central disk galaxy during the evolution of the system, showing in general $\alpha>70^\circ$, and an average angle of $\alpha\sim80^\circ$ at early times. At $T_{\rm uni}\approx 9.4$ Gyr, the persistent plane is mostly perpendicular to the central galaxy's disk. 
We recall that PDEVA-5004's  disk spin direction  changes very little along the analyzed period and therefore the variations found in  $\alpha(\vec{n}_{\rm KPP},\vec{J}_{\rm disk})$ are   mostly due to  fluctuations in the orientation of PDEVA-5004's main $\vec{n}_{\rm KPP}$ vector.
PDEVA-5004's second plane (green, third column in Figure~\ref{fig:kineplane-ori})  is, interestingly, also remarkably perpendicular to the central disk galaxy at early times. 
However, after $T_{\rm uni}\sim10$ Gyr, 
the perpendicularity is lost and the angle formed with the disk is reduced to $\sim45^\circ$\footnote{This change could be triggered by satellite accretion and multiple pericentre coincidences, taking place at similar times. However, this issue is beyond the scope of this work.}.

It is noteworthy that both PDEVA-5004's persistent planes are perpendicular to the central galaxy's disk for $T_{\rm uni}<10$ Gyr. 
Interestingly,
the two planes are also exactly perpendicular to each other at early times, until $T_{\rm uni}\approx8.5$ Gyr.
 Afterwards, the angle between them decreases drastically (due to the change in direction of the second plane), but then 
at T$_{\rm uni}\sim10$ Gyr it
 starts increasing steadily, reaching an angle of $\sim 70^\circ$ by $z=0$.

To finish, let us add that no remarkable mutual effects between the  spin vector of the central galaxy and the orbital pole directions of satellites (or the normal direction to the persistent plane) have been appreciated, suggesting kinematically-coherent satellites are least affected by local, secular effects produced by disk changes and may have a cosmological origin.
 This is consistent with the good conservation of the angular momentum $\vec{J}_{\rm orb}$ showed by many satellites. 
We note however that their interdependence is unclear from this analysis alone: a study of the total gravitational potential, and of the contribution of the central galaxy's disk to it,  would be necessary for a deeper understanding of the role played by the disk in possibly `dragging' the plane of satellites. 
In principle, satellites will only feel the disk's influence when passing nearby, although the timescale of these close passages may also be important. Such a detailed study is beyond the scope of this work.




\section{Properties of kinematically-coherent satellites} \label{sec:PropKineSat}

\begin{figure*}
\includegraphics[width=0.49\linewidth]{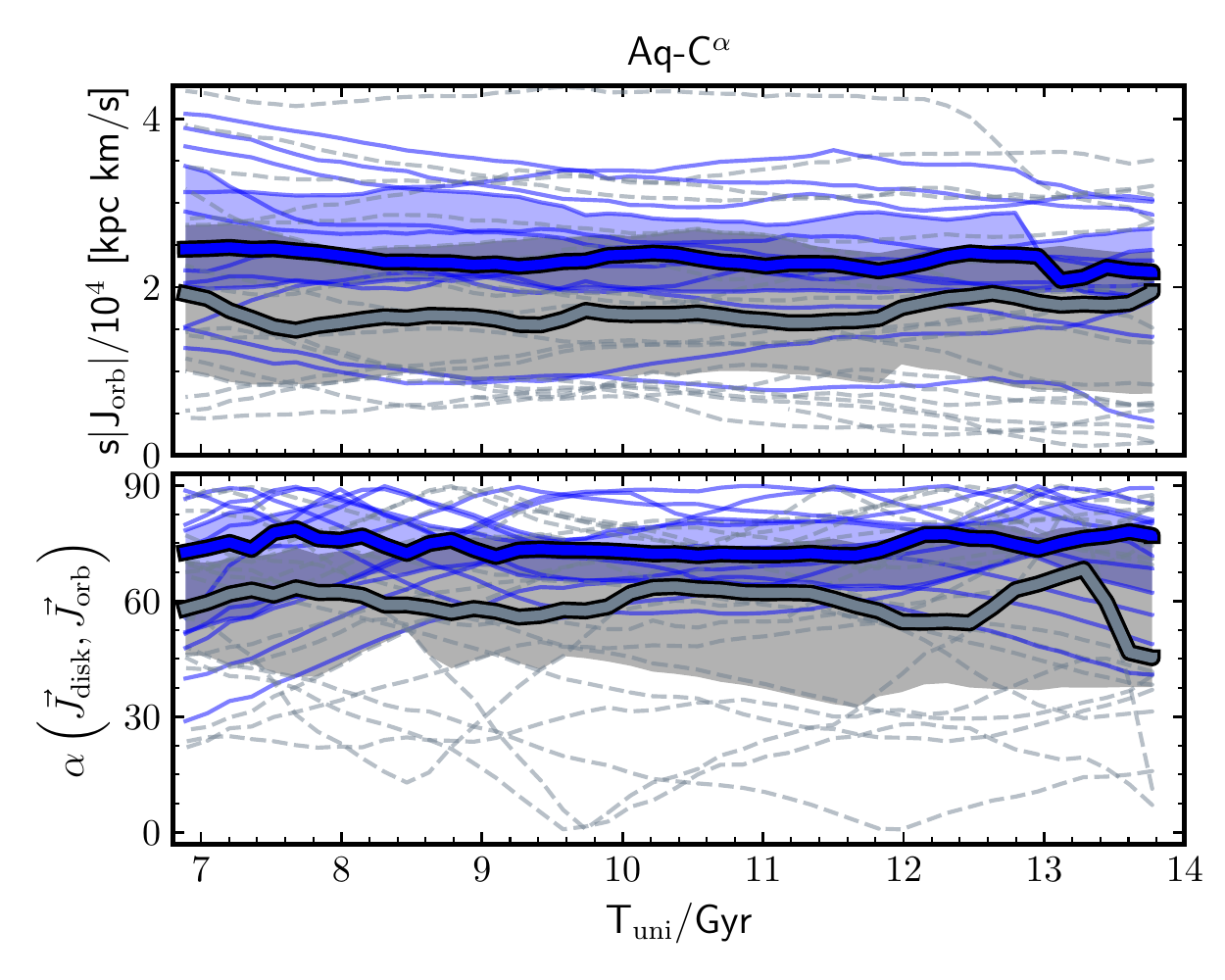}
\includegraphics[width=0.49\linewidth]{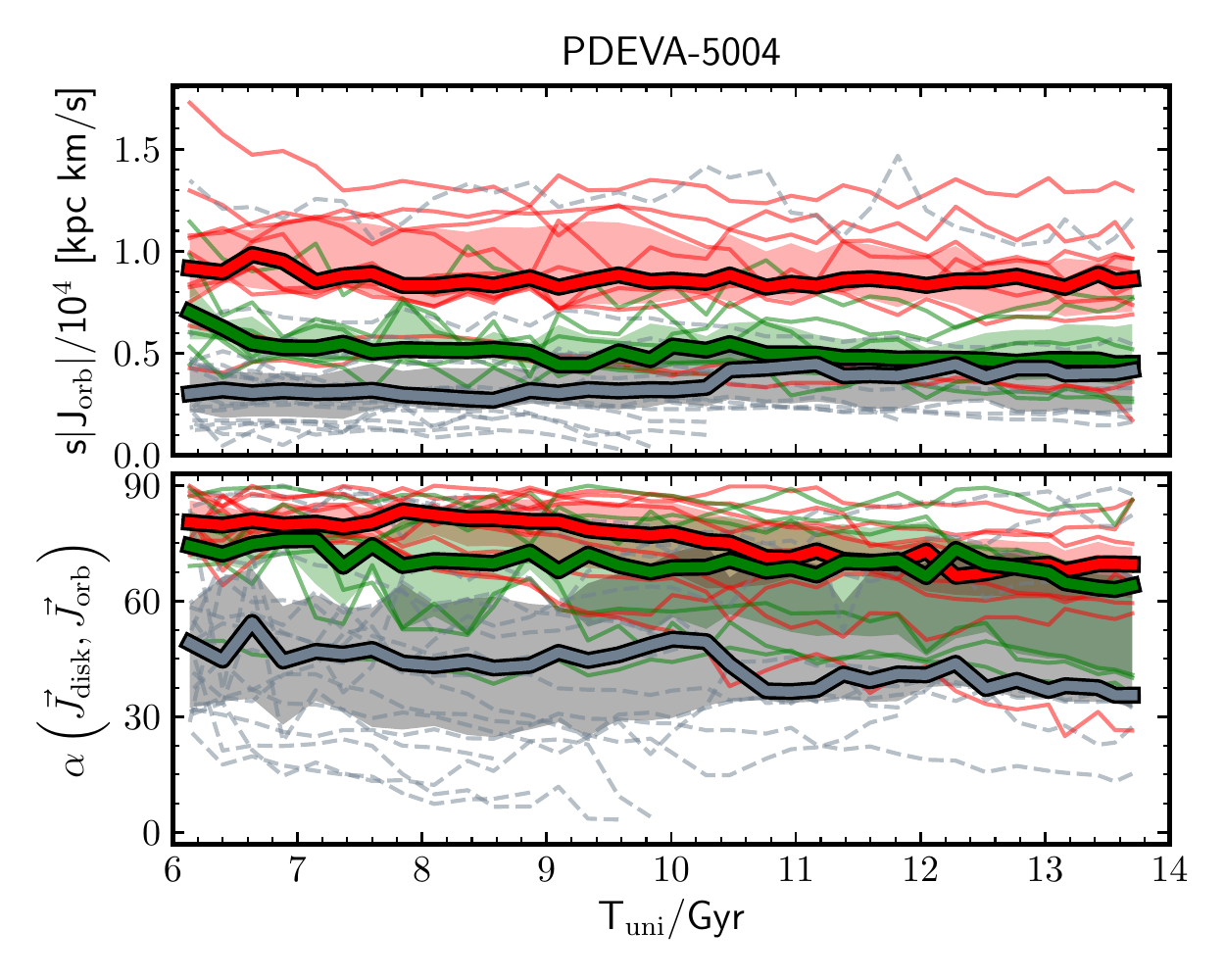}
\caption{Time evolution of properties of the kinematically-coherent satellites identified in each simulation.
Left: Aq-C$^\alpha$, right: PDEVA-5004.
Top panels show the magnitude of the specific orbital angular momentum  $sJ_{\rm orb}$,
and  bottom panels show
the angle formed between the spin vector of the central disk galaxy and  the orbital pole of satellites, $\alpha\left(\vec{J}_{\rm disk}, \vec{J}_{\rm orb} \right)$.
Thin lines show results for individual satellites, shown in color for kinematically-coherent satellites (blue for the main group in Aq-C$^\alpha$, red and green for the main and second groups in PDEVA-5004, respectively), or in dashed-gray for non-kinematically-coherent ones. 
Following the same color scheme, a thick line with shade indicates the median and the 25-75 percentile range,
calculated at each timestep.}
\label{fig:propkine}
\end{figure*}

In order to know if the kinematically-coherent   satellites  represent a family of satellites with any special common characteristics, differentiating them from the non-kinematically-coherent ones, we have analyzed their properties as a group. Particularly, we focus on:
i) the magnitude of the satellite specific orbital angular momentum,  $sJ_{\rm orb}$, along time, 
ii) the angle formed between the spin vector of the central disk galaxy and the satellite orbital angular momentum vector, $\alpha\left(\vec{J}_{\rm disk}, \vec{J}_{\rm orb} \right)$, along time,
iii) the pericenter  distance of satellites to the central disk galaxy, $dp$, along time,
and iv) the satellite mass.
Figure~\ref{fig:propkine} shows the time evolution of  $sJ_{\rm orb}$ and $\alpha\left(\vec{J}_{\rm disk}, \vec{J}_{\rm orb} \right)$,
for both Aq-C$^\alpha$ (left panel) and PDEVA-5004 (right panel) satellite populations.

Individual lines in the top panels of Figure~\ref{fig:propkine} show that, 
in the slow phase of mass assembly, $sJ_{\rm orb}$ is  overall conserved in many cases\footnote{It is worth noting that the overall values of $s{J}_{\rm orb}$ are factors of $\sim 2$ larger for satellites in  Aq-C$^\alpha$ than in PDEVA-5004. The reason is that the halo mass in the former system is a factor of $\sim4$ higher than the latter halo mass. High mass halos collapse later than low mass ones, and, consequently,   acquire higher angular momentum, as predicted by Tidal Torque Theory \citep[see e.g.][]{Lopez2019}.}.
Focusing on the main groups of kinematically-coherent satellites of each simulation (blue and red lines),
we can see that they
present overall larger
$sJ_{\rm orb}$
 than the rest of satellites (dashed-gray lines), shown by clearly separated median values 
across time
(except at the very end of the period analyzed for Aq-C$^\alpha$). 
Kinematically-coherent satellites also tend to orbit more perpendicularly to the central disk galaxy than the rest, with larger values of $\alpha\left(\vec{J}_{\rm disk}, \vec{J}_{\rm orb} \right)$ (see bottom panels).

Interestingly, KPP satellites also show significantly larger pericentric distances $dp$ than the rest of satellites:
the main groups in Aq-C$^\alpha$ and PDEVA-5004 show a median $dp$  of
$\sim78$ ($46$) kpc in Aq-C$^\alpha$ (PDEVA-5004), versus the  $\sim38$ ($12$) kpc shown by non-kinematically-coherent satellites.
These results imply KPP satellites tend to be less affected by disk tidal effects, a factor favoring the long-term conservation of satellite orbital angular momentum.

To confirm quantitatively that the subgroups of kinematically-coherent satellites are different to  the rest of satellites in terms of $sJ_{\rm orb}$, $\alpha\left(\vec{J}_{\rm disk}, \vec{J}_{\rm orb} \right)$ and $dp$, we have performed statistical Kolmogorov-Smirnov tests (KS-tests) between the different satellite subsamples.
To perform these KS-tests we have assumed each satellite contributes to the sample as many times as timesteps in which it participates. As an example, if a subsample presents $N_{\rm sat}=10$ and there are $N_{\rm timesteps}=40$ timesteps in total, we perform the KS-test for that sample with $10\times 40$ elements. We have done these tests for the periods analyzed
after halo virialization, i.e. from T$_{\rm vir}$ to T$_{\rm z=0}$, and taking as subsamples the main group of kinematically-coherent satellites, 
on the one hand,
and the rest of satellites of the system, on the other hand.
We find  that the null hypothesis (i.e., that the two subsamples are drawn from the same underlying population) can be rejected to a higher than $99.9\%$ confidence level, 
as far as 
$sJ_{\rm orb}$,
$\alpha\left(\vec{J}_{\rm disk}, \vec{J}_{\rm orb} \right)$ and $dp$  are concerned.

We focus now on the second group of kinematically-coherent satellites in PDEVA-5004 (green in Figure~\ref{fig:propkine}).
Concerning the parameters studied, KS-tests confirm that this   group   is statistically different to the main group (red). It is also different to the set of satellites that are not associated to either kinematic group (gray), despite the distributions of both samples showing similar medians during some time periods. 
In this case, kinematically-coherent satellites have 
$sJ_{\rm orb}$ lower than those on the main plane,
but still slightly larger than for non-kinematically coherent satellites.
These KPP2 satellites also orbit reaching closer pericentric distances to the central galaxy than those of satellites on the main KPP1 plane. Specifically, we find  a median $dp$ of $\sim 22$ kpc: only slightly larger than the value for non-kinematically-coherent satellites.
On the other hand, the angle formed between the orbital plane of satellites  and the galactic disk is large
($\sim 70^\circ$); slighty lower but comparable to values found for the main group, 
evidencing that persistent satellites are not on orbits co-planar with the central disk galaxy.

Finally, in addition to the time evolution of the parameters above, we have also studied the differences between
satellites in KPPs and outside them
 in terms of total baryonic mass. 
We compare the masses of satellites 
at a fixed time of $z=0.5$ for both simulations.
In this case, a KS-test shows that 
it is not possible to reject the null hypothesis. 
This result agrees with the indistinguishable luminosities and masses observed for M31's on- and off-plane satellites \citep{Collins15}.
It is also consistent with results emphasized in Paper II, where it was found that baryonic mass was not a property driving a higher probability of belonging to high-quality positional planes of satellites. 
This finding is relevant as it shows that the relatively limited mass resolution that current simulations can reach (and therefore limited range of masses found for simulated satellite populations),  does not seem to introduce a bias when studying planes of satellites using simulations and comparing the results to observational data (where the mass range spanned may be larger).

%
%

\section{The late capture of a massive dwarf galaxy and its sub-satellites in Aq-C$^{\alpha}$ }
\label{sec:late-capture}

\begin{figure*}
\centering
\includegraphics[width=0.42\linewidth]{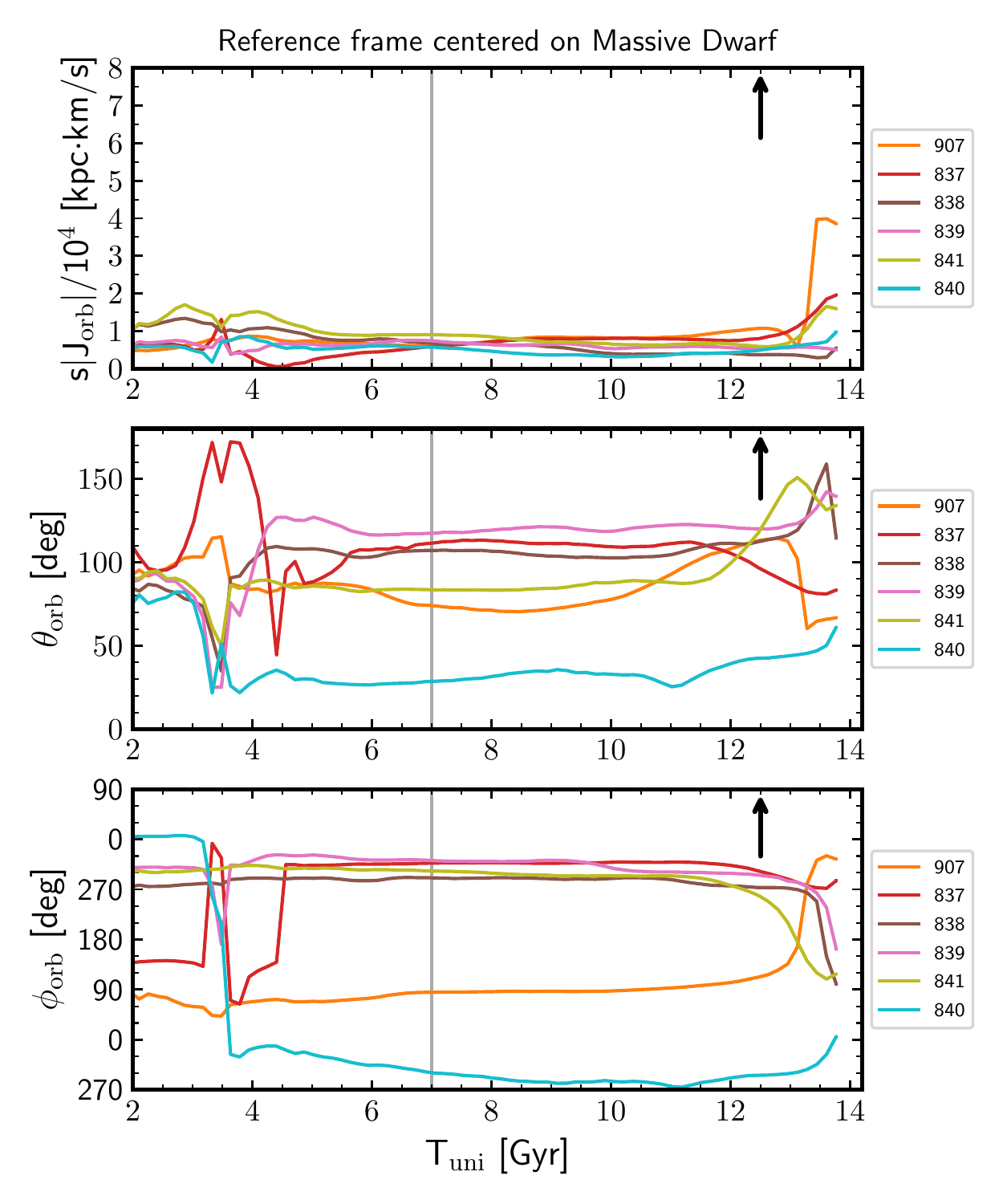}
\includegraphics[width=0.42\linewidth]{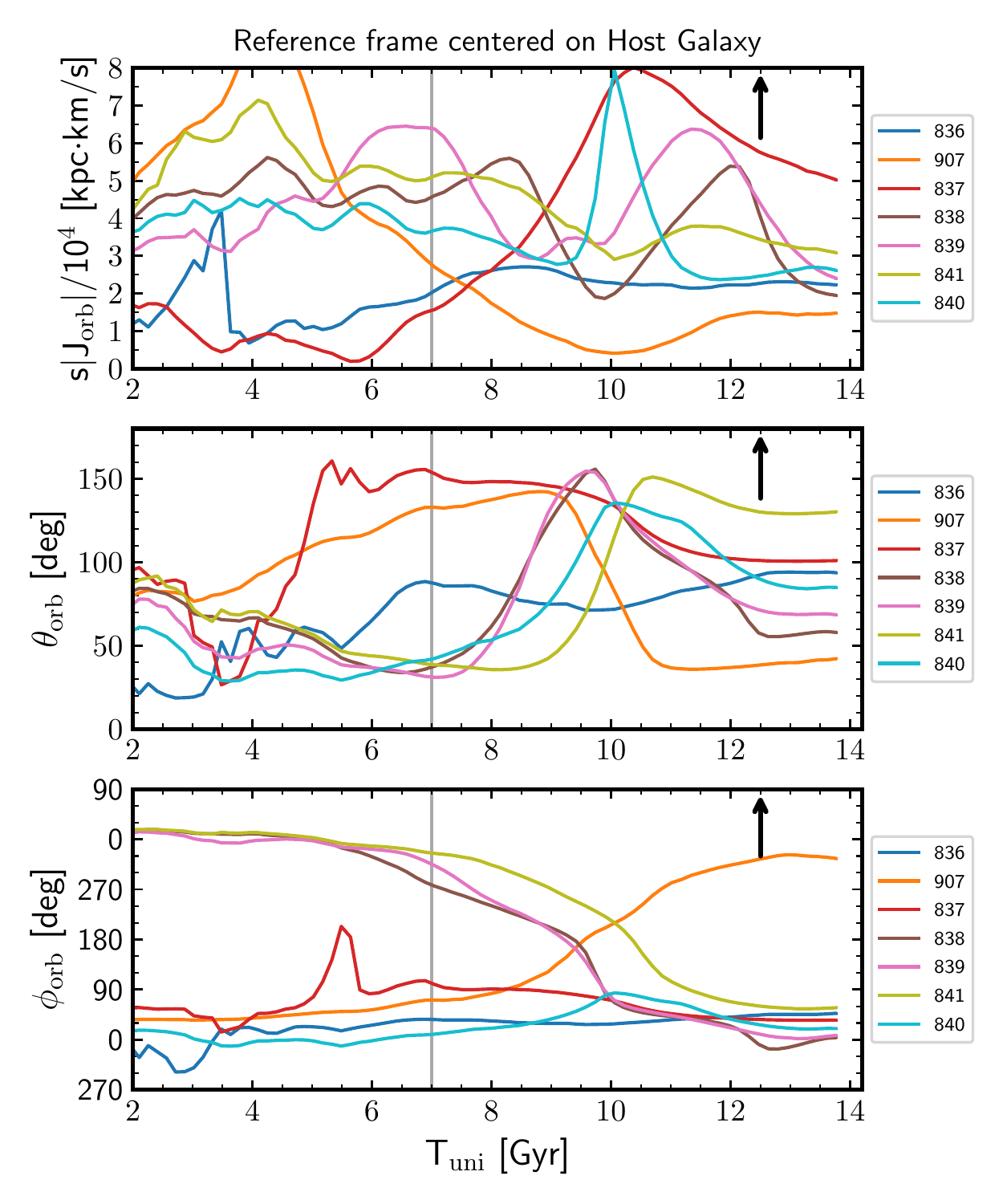}\\
\includegraphics[width=0.6\linewidth]{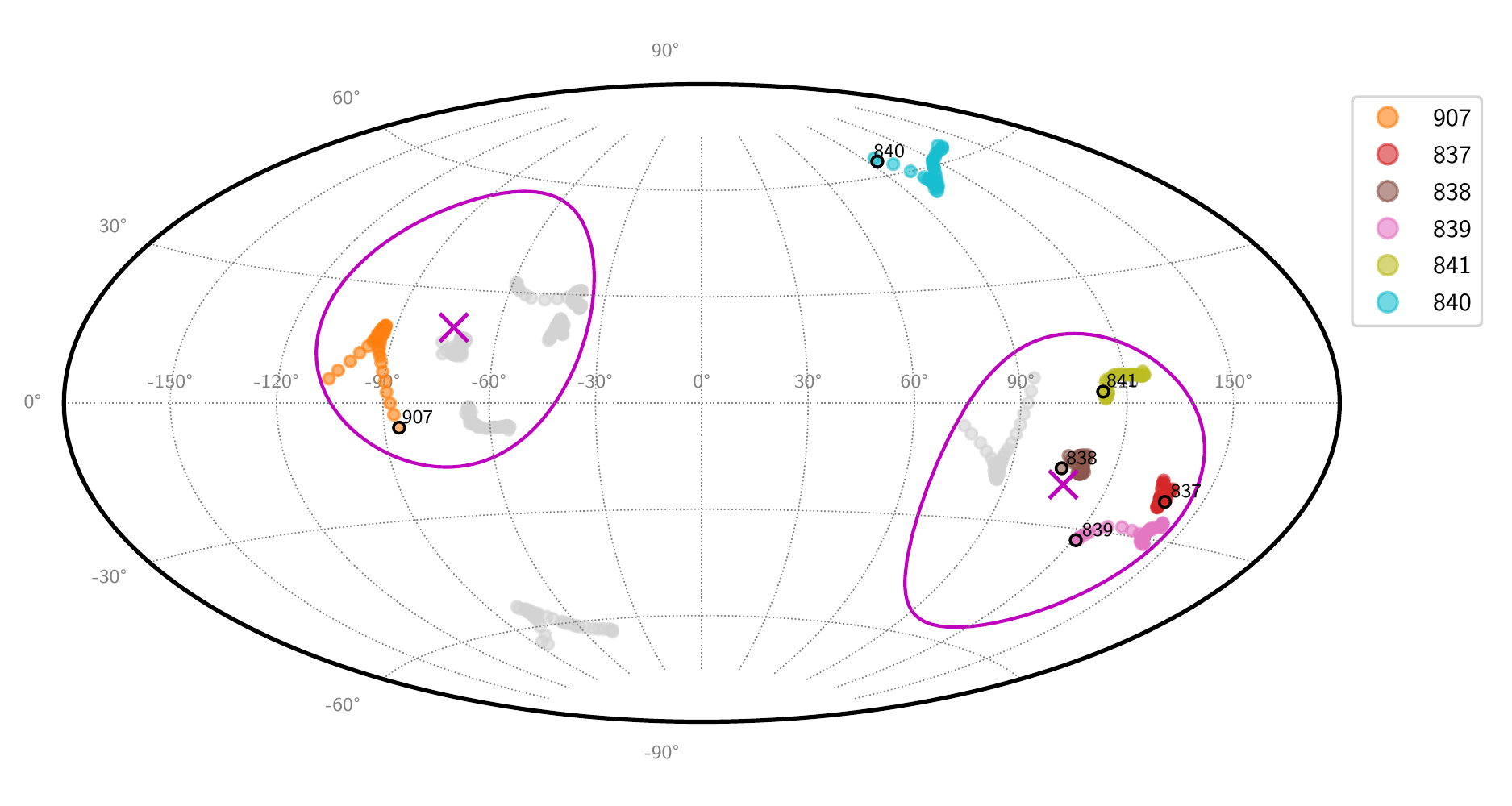}
\caption{The late capture process of a massive dwarf (MD)  with its own satellite system in Aq-C$^{\alpha}$. The three upper  rows give  the components of the $\vec{J}_{\rm orb}$ vector of each sub-satellite, as a function of time, relative to the massive dwarf (left panels) and to the  main galaxy
 (right panels). 
 The evolution of  the MD's  $\vec{J}_{\rm orb}$ vector relative to the main galaxy  is  also given in the right panels (ID \#836).
Vertical arrows  mark the capture time (T$_{\rm uni}\approx 12.5$ Gyr), when sub-satellites become bound to the host, i.e., the beginning of the time interval along which orbital angular momentum relative to the MD is transferred to the main galaxy.
The bottom plot shows an Aitoff projection of the MD's sub-satellite orbital poles, centered on the MD, along  the T$_{\rm uni}$ interval between 6.0 and 11.1 Gyr, before this system is captured by the host. The corresponding $\vec{J}_{\rm stack}$ axes are marked with `x' crosses, and 
circles mark the co-orbitation region.
Gray tracks indicate the diametrically opposite projections of each orbital pole, as we consider these as axial vectors in our method (see Section~\ref{sec:methodsims}).
}
\label{fig:jsat_subsate}
\end{figure*}
 
The Aq-C$^{\alpha}$  system suffers a potentially disturbing dynamical event beginning by  T$_{\rm uni} \sim $ 11.5 Gyr
due to the late infall of a satellite group.
To elucidate whether or not this event affects Aq-C$^{\alpha}$'s  KPP, 
we had considered different `stacking' time intervals in our analysis in search for KPPs:
i) the entire slow-phase assembly period from T$_{\rm vir}$ to T$_{\rm z=0}$, 
and, ii), from T$_{\rm vir}$ to T$_{\rm uni}$ = 11.5 Gyr.
As mentioned in subsection \ref{sec:methodsims}, the directions of the $\vec{J}_{\rm stack}$ axes found in each case show  no remarkable  differences.
We had concluded that this dynamical event has no perturbing effects on Aq-C$^{\alpha}$'s  KPP.

We now  study some details of the dynamical event. Is this a merger or a capture? Are they simple or multiple events?
An analysis of the Aq-C$^{\alpha}$  system beyond T$_{\rm uni}$ = 11.5 Gyr through the Subfind halo finder  and  completed with the
recently developed  visualization device  IRHYS-2, allowed us to identify 
the capture of a massive dwarf, MD, galaxy.
It has a specific orbital angular momentum relative to the main galaxy with a magnitude of  $sJ_{\rm MD, main} \sim 2.2 \times 10^4$ kpc km/sec,  conserved since T$_{\rm uni} \sim7$   Gyr 
(see Figure~\ref{fig:jsat_subsate}, right panels, object with ID number \#836).
In addition, 
the MD's orbital  pole with respect to the main galaxy  is conserved from T$_{\rm uni} \sim7$   Gyr onwards.
At $z=0$,  the  MD  is at a distance  $D_{\rm MD, g}$ = 65 kpc from the main galaxy, with the  $\vec{J}_{\rm disk, MD}$ and $\vec{J}_{\rm disk, main}$ vectors roughly parallel to each other.

This LMC-like galaxy carries 6 sub-satellites (with ID numbers \#907, 837, 838, 839, 840, 841), 
identified according to the criteria given in Section \ref{sec:satIdent}.

   Their orbital angular momentum relative to the MD galaxy is conserved
from high $z$   
    up to T$_{\rm uni} \sim$ 12  Gyr. After this time, angular momentum transfer to the main galaxy begins, see Figure~\ref{fig:jsat_subsate}. 
   The bottom panel shows an Aitoff projection of  the MD's satellite orbital poles, centered on the MD, along  the T$_{\rm uni}$ interval between 6.0 and 11.1 Gyr, before this system is captured by the host galaxy. 
    This Aitoff projection shows that 
5  out of 6 of the sub-satellite orbital  poles are clustered   prior to incorporation:  IDs \#837, 838, 839 and 841, showing rotation in the same sense, and satellite  \#907, which is rotating in the opposite sense.
Remarkably, the MD galaxy not only carries its own satellite system, but, in addition, this system 
is kinematically structured,
orbiting in a common plane 
 with a high co- versus counter-rotating satellite number ratio (4/1).

Satellites of  the MD  galaxy system gradually become members of the main galaxy  satellite system.
The orbital poles  of the 6 MD satellites relative to the main galaxy  are roughly conserved from T$_{\rm uni} \sim$ 12  Gyr onwards
 (see Figure~\ref{fig:jsat_subsate}, right panels).
One pericentric passage happens
for satellites \#907 (at 13.3 Gyr) and \#841 (at 13.5 Gyr). For the other 4 satellites, no pericenter relative to the main galaxy has been reached yet before $z=0$. The same is true for the MD galaxy itself.

 As for  clustering 
in orbital pole space
 at low $z$, the poles of these  7  newly captured satellites  span a linear  arc in the spherical projection centered on the main galaxy.
By $z \sim 0$, 3 out of the 7  newly captured satellites  (IDs \#836(MD), 837 and 841) are
located  within the left co-orbitation circle of Figure~\ref{fig:Aitoff_AqC_ALL}A
(note that the MD's orbital pole relative to the host galaxy clusters with the original KPP bundle 
 since times much earlier to its capture).
This adds 3 satellites
to Aq-C$^{\alpha}$'s KPP,
 co-rotating together with those printed  red  in Table~\ref{table:summary-discard},
increasing the  ratio of  
co- over counter-rotating  satellites from 
7/6 to 10/6.\footnote{This fraction is consistent with the lower-bound ratio of 5/3 found for the MW in Paper I. However, the wide range of possible ratio values we have obtained prevents us from reaching  any strong  conclusion about this point.}

These results show that the late capture of an LMC-like satellite group could have implications in the appearance of highly structured kinematic planes at low $z$, potentially enhancing asymmetry in the number of satellites
rotating in one sense versus those rotating in the contrary.

%
%

\section{Discussion: KPPs versus positional planes}
\label{sec:KPPvsPosPla}
Different works have claimed that 
the highest-quality
positional planes 
one can identify in cosmological simulations
are  unstable entities, meaning that they lose or change an important fraction of their satellite members in a short timescale
(e.g. \citealt{Gillet15,Buck16}, Paper II; see Section~\ref{sec:intro}).

Positional planes in Paper II had been selected as the highest quality ones (i.e., lowest $c/a$) among those encompassing a fixed fraction of satellites of the whole sample, at a given time (see figure~9 in Paper II). This selection criterion has as a consequence that the IDs of satellites in these positional structures change from one time-output of the simulation to the following, imprinting fluctuations to the 
curves representing the time evolution of 
$\alpha(\vec{n},\vec{J}_{\rm disk})$
(the angle formed between the normal vector to the plane, $\vec{n}_{\rm pos\%}$, and the host galaxy's axis).
As the number of possible combinations of satellites increases when the fixed fraction of satellites decreases, these fluctuations become more frequent and of higher intensity as the fraction considered decreases.

\begin{figure}
\centering
\includegraphics[width=\linewidth]{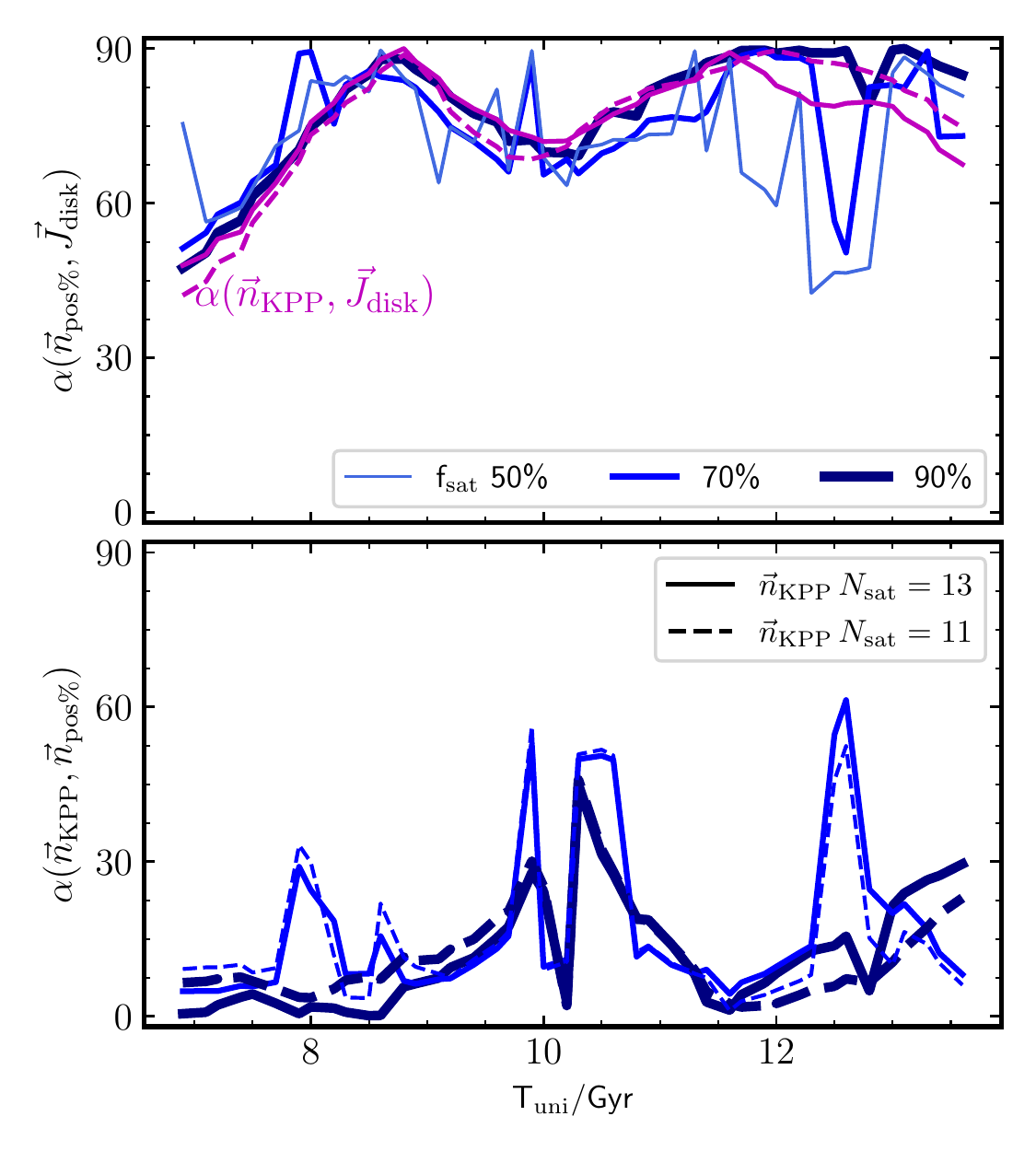}
\caption{
 Comparison of the directions of the   normals to the KPPs, ($\vec{n}_{\rm KPP}$, lower  block of panels in Figure~\ref{fig:kineplane-ori})  
with the normals to  positional planes ($\vec{n}_{\rm pos\%}$, identified in Paper II),   
 for the  Aq-C$^\alpha$ system.  The top  panels show the angles  both normals form with the host disk axis $\vec{J}_{\rm disk}$ along cosmic time. Blue lines refer to the  best  positional planes involving a given fixed fraction of the total number of satellites. The different shades of these lines stand for these different fractions,  as specified in  the legends.  
The magenta lines refer to angles involving the   normals to the KPPs,
with number of satellites as specified in the legends.
In the bottom panel the evolution of the  angle formed between the  kinematic normals and the positional  normals is drawn. 
} 
\label{fig:PosKKP-AqC-Comp}
\end{figure}

The angle formed by 
$\vec{J}_{\rm disk}$ with the 
normal vector to
KPPs is given in the bottom panels of Figure~\ref{fig:kineplane-ori} here.
In this case, as already mentioned, the evolution curves are rather smooth and do not depend that much on the number of satellites involved, a manifestation of the robustness of KPPs as  kinematic entities. 
In the top panel of Figure~\ref{fig:PosKKP-AqC-Comp} we compare 
the evolution of the angle formed with $\vec{J}_{\rm disk}$ 
by the normals of both the
kinematic planes (magenta curves) and the positional planes as described above (blue curves), for the Aq-C$^\alpha$ sample. 
We see that the angles formed by kinematic and positional planes evolve in unison,
 showing their minima and  maxima at roughly the same times, and with similar intensity.

This is a necessary condition for the alignment of 
positional and kinematic
normals to occur. We explore such an alignment in the bottom panel of Figure~\ref{fig:PosKKP-AqC-Comp}.

The best alignments 
between $\vec{n}_{\rm pos\%}$ and $\vec{n}_{\rm KPP}$
occur when the fraction of satellites involved in positional planes is of a 90\%
(depicted by thick dark blue lines  in Figure~\ref{fig:PosKKP-AqC-Comp}).
The number of satellites with which we track KPPs  has comparatively a lower impact. 
We see that, in 
the 90\% case,
normals are aligned within an aperture better than  10$^\circ$ for roughly half the time interval after T$_{\rm vir}$.
They are aligned to better than 15$^\circ$ along the whole interval considered, except for T$_{\rm uni}$ between 
9.5 - 11
Gyr and after 13 Gyr. 
The alignment is worse than 30$^\circ$ around T$_{\rm uni}$ = 10 Gyr.
In Paper II (see their figures~6 and 9) we argued that the quality of positional planes in the Aq-C$^\alpha$ set is low along  this same time interval. 
On the other hand, the quality of KPPs as planar structures, including orientation changes,   worsens somehow around 10 Gyr as well. 
A plot of the satellite radial distances to the host center indicates that there is a pericenter accumulation at this epoch,
 possibly causing these disturbances to positional as well as to kinematic planes. Assessing their causes is beyond the scope of this paper though. 

As said above, when a lower fraction of satellites is used to track positional planes instead, fluctuations become more and more important.
With a fraction of 70\%, alignments slightly worsen around T$_{\rm uni}$ = 8 Gyr, where, again, the quality of positional planes 
diminishes, see figure 6 in Paper II. 
However, the overall alignment with $\vec{n}_{\rm KPP}$ remains  robust qualitatively.

This alignment between normals to KPPs 
and normals to the best quality positional planes, 
implies that high-quality positional and kinematically-identified  planes share the same  configurations in positions at fixed times.
KPPs conform a kind of skeleton, persistent along time, to which other satellites add in positions during given limited time intervals.
However, the latter  are not in coherent  kinematic co-orbitation with the former; at least not during  long time intervals.
Consequently, they become lost to the positional configuration after a timescale of the order of the time it takes 
for each of them to cross the plane.

When the satellite system under consideration has two kinematic planes with similar collimation and intensity, the fluctuations 
in properties of positional planes described at the beginning of this section become even more important, making it more difficult to implement
the analysis we have just made for the Aq-C$^\alpha$ satellite system.
This is the case for the PDEVA-5004 system after T$_{\rm uni} \sim $ 10  Gyr. In this case, KPP satellite members of the two
kinematic entities are mixed in figure 9 of Paper II, making it difficult to reach any firm conclusion by 
applying the protocol described above.

\section{Summary and conclusions}\label{sec:conclu}
\subsection{Summary}
In this paper we analyze two zoom-in hydro-simulations of disk galaxy formation run in a  $\Lambda$CDM cosmological context, 
Aq-C$^\alpha$ and PDEVA-5004, 
that make use of different initial conditions, codes and subgrid physics. 
We identify, in each of them, a skeleton of kinematically-coherent, thin planes of satellites, whose member identities persist  along $\sim 7 $ Gyr of cosmic evolution  at least: the so-called kinematically-coherent persistent planes, or KPPs. 
By kinematic coherence we mean satellite motions in such a way that their orbital poles are conserved and 
are clustered for a long period of time, rotating within a common orbital plane in one  sense or the opposite.
These identifications are  a step forward towards understanding when planes of satellites are set, and whether they can possibly persist along long periods of time or not. We do not intend a detailed comparation to the MW data. 
The issue of satellite plane persistence is relevant per se, and more importantly, because it could be the consequence of KPP satellites having gained their common dynamics at early times, in unison with the Cosmic Web local dynamics (G\'amez-Mar\'{i}n  et al., in preparation).
By identifying KPPs, we overcome the drawbacks inherent to methods using three-dimensional satellite   positions only,
which detect good quality spatial planar structures, but where only a fraction of satellites are kinematically-coherent.

The orbital poles of satellites with good enough orbital angular momentum conservation have been stacked together along time intervals when no relevant disturbing dynamical events occur
(from halo collapse  -or virialization time- to  $z=0$). Their projection on the sphere has been scanned to look for directions around which  the orbital poles accumulate. 
This protocol has been applied here to the outputs of the two  simulations mentioned above,  where a massive disk galaxy, radially extended, with a thin disk made of stars and gas,  has been identified in each of them.

The analysis returns the directions of stable-in-time axes, $\vec{J}_{\rm stack}$,  around which the number of co-orbiting satellites
is maximum: we identify one for Aq-C$^\alpha$ and two for PDEVA-5004.
KPP satellite members, corresponding to a given axis $\vec{J}_{\rm stack}$,  
 have been selected as  
 those whose orbital poles are  within an angular distance of $\alpha_{\rm co-orbit}=36.87^\circ$ from $\vec{J}_{\rm stack}$ at least for  50\% of the simulation outputs. 
It turns out that the satellite IDs of KPPs this protocol returns are largely the same as those returned by the 
so-called `3-$\vec{J}_{\rm orb}$-barycenter method', see Paper I. This is a verification of the robustness of our results.

The groups of KPP satellites are conformed by a relatively high fraction of the total number of satellites in the samples studied in this paper.
The fraction is higher in the case of the  Aq-C$^\alpha$ KPP ($\sim40\%$), and lower in each of the two PDEVA-5004 KPPs,
where, when the contributions from the two KPPs  are summed up, a fraction as high as an $\sim 80\% $
of kinematically-structured satellites is found during some time intervals. 
Within each KPP, the fraction of satellites captured at high $z$ 
 rotating in one sense or the opposite generally tends to be of a   $\sim 50\%$ 
in either simulation.
This situation could be
changed by the late capture of a massive dwarf
with its own satellite system, see below.

These numbers on co-orbitation are  consistent with the fraction of kinematically-coherent satellites found in the MW system given the current data uncertainties ($\sim 40\%$, see Paper I). However, when the sense of rotation is taken into account, the \% of coherent rotation in the MW is higher (a minimum of 5/3 considering measurement errors, see Paper I).

Planes fitted to the positions of each of  these groups of kinematic satellites
across time show that
KPPs present  overall good qualities  in terms of a Tensor of Inertia (ToI)  analysis, and improve  during given time intervals.

The angle formed by the  normal to the KPPs and the central galaxy's spin vector,
$\alpha(\vec{n}_{\rm KPP}, \vec{J}_{\rm disk})$,
 changes with time, its evolution reflecting the combined effect of disk `flipping' (important in Aq-C$^\alpha$) and the change of orientation of the KPP (not very relevant in either simulation). 
Curiously in both simulations, the KPPs 
show an approximately polar orientation with respect to their central disk galaxies over long time periods.

Our results seem to suggest that, contrary to positional planes,  the  KPPs are  little affected by local gravitational forces induced by changes in  disk orientation or instabilities. This could imply that they may have  a cosmological origin.
 We note, however,  that some  changes 
 in the KPPs of the PDEVA-5004 system occur at a moment of multiple satellite pericenter passages 
 and could be due to disk torques on the closest satellites. Deepening in these ideas is outside the scope of this paper.

We have compared the time evolution of  the $\alpha(\vec{n}_{\rm KPP}, \vec{J}_{\rm disk})$ angle, on the one hand,  with that of  the  
angle formed by $\vec{J}_{\rm disk}$ and 
the normal to the best-quality positional plane encompasing a fixed fraction of satellites,
on the other hand. In the case of the Aq-C$^{\alpha}$ simulation, these evolutions follow each other. 
Moreover, the normals to KPPs and those to the positional planes just described, are roughly aligned, except in a time interval  when an accumulation of satellite pericenters occurs not far from the host galaxy center.
This result implies that KPPs and the best-quality positional planes involving fixed satellite fractions, are the same configurations in three dimensions in the case of Aq-C$^{\alpha}$.

As for the PDEVA-5004 system, both their positional and kinematic configurations (with two KPPs) are complex and they mix up, making it difficult to reach  any strong conclusion  by applying the same protocol as  in the previous case.

We have looked for possible indications of a statistically distinguishable behaviour between  KPP satellites  and non-KPP ones. 
By means of Kolmogorov-Smirnov (KS)-tests, we find that the former are a distinguishable subsample of satellites as compared to the latter regarding the magnitude of their specific orbital angular momentum $sJ_{\rm orb}$, the angles between satellite orbital poles and the host galaxy's spin,  and the pericentric distances to the center of the host galaxy.
These  magnitudes, whose medians show a time-dependence,  are statistically  larger for KPP members than for satellites outside these kinematic structures.
These differences are less marked in the case of the second KPP in PDEVA-5004.

Concerning total satellite baryonic mass, KS-tests manifest that the differences are not statistically significant.
In other words, at least in these two simulations and within the mass range spanned by satellites,
the baryonic mass is not a property that determines whether a satellite belongs or not to the corresponding persistent plane.
This is an important result given the limited satellite mass range  allowed by  current computational possibilities.

\subsection{Some comparisons to previous work}

To our knowledge this is the first work highlighting the finding and detailed characteristics of
$\Lambda$CDM-simulated
 planes of kinematically-coherent satellites orbiting around disk galaxies, 
which are persistent since at least halo virialization.

In a similar fashion to this work, \citet{Garaldi18} also studied satellite orbital pole clustering, identifying  highly-populated $z=0$ planes including a high fraction of co-orbiting satellites  
in two out of the  4 galactic systems they studied,
just in those haloes containing well-developed spiral galaxies. 
In one case the satellites maintain clustered orbital poles for a long period of time,
with $\sim1/3$ rotating in one sense and $2/3$ in the opposite,
while in the other case all the satellites rotate in the same sense on  a kinematic plane set at recent times. 
Their $c/a$ results are in general compatible to those found here for persistent planes in Aq-C$^\alpha$ and PDEVA-5004. However  no systematic searches of co-orbiting satellite groups, studies of satellite-membership persistence, or supplementary plane characteristics (e.g., oblateness, orientation)  were presented in their work, preventing a proper comparison with our results.

On their part, \citet{Samuel2021} report that, even if rare, they do find, within  the $z=0.2-0.0$ period they analyze,  planes as thin and as kinematically coherent as the MW one 
including the high fraction of co-rotating satellites,
 favoured by the presence of LMC-like satellites 
undergoing pericentric passages.

In line with these ideas, 
we have studied  Aq-C$^\alpha$'s late
  capture  at T$_{\rm uni} \sim 12$ Gyr
 of a massive dwarf galaxy endowed with its own sub-satellite system (6 satellites, under the selection criterion used in this work).
While bound to the massive dwarf (prior to the  capture of the system), 5 out of the 6 sub-satellites co-orbit within  a common plane (and 4 of them rotate in the same sense). 
Once bound to the host galaxy, 3  of these  new satellites have been found to 
co-rotate within the original KPP plane, increasing the co- versus counter-rotation ratio at $z=0$  
(i.e., adding 3 red-colored satellites to results in  Table \ref{table:summary-discard}).
These results go in the sense of  \citet{Samuel2021}'s findings.

\subsection{Conclusions}
Summing up, these are the  main conclusions of this paper:
\begin{itemize}[noitemsep, nosep, leftmargin=*]
\item In each of the two simulations analyzed here,   specific fixed  sets of satellites have been found that are kinematically-coherent  across the slow phase of mass assembly, 
forming a kinematically-persistent plane of satellites (one in Aq-C$^{\alpha}$ with  13 satellites, two in PDEVA-5004 with 10 and 7 members, respectively). This is a consequence of an adequately good angular momenta conservation. 
KPPs show  similar common patterns in both simulations, in spite of their different subgrid physics, feedback implementation, etc.

\item   These time-persistent kinematically-coherent sets of satellites form the (stable-in-time) skeleton of satellite planes  around galaxies, preventing them from being  washed  out on short timescales. 
We have proven that good quality positional planes of satellites and KPPs share the same space configuration in three dimensions.
Indeed, positionally-detected  satellite planes consist of the persistent skeleton, plus transient satellites, that happen to cross the KPP at the time of observation. The latter are positional plane members during one crossing timescale, being replaced later on by other transient satellites. 
In this way, positional satellite planes are long-lasting configurations, although 
unstable ones.

\item Satellites in KPPs and outside these structures are statistically distinguishable, concerning specific angular momentum and  pericentric distances (higher and larger in the former than in the latter, respectively). Satellites in KPPs 
tend to move on nearer-to-polar orbits too. 
However,  mass is not a satellite property that determines its membership to a KPP, within the limits of this work.

\item Our results suggest that a recent accretion of a massive dwarf carrying its own satellite system represents a possible low-redshift channel enhancing KPPs.
Indeed, while it is not a necessary condition to form thin planes of long-lasting kinematically-coherent satellites, it might contribute, or might even be necessary, to enhance the fraction of satellites rotating in the same sense in KPPs.

\end{itemize}
 The persistent-across-time character of the kinematically-coherent planes of satellites suggests that they  have been set at high redshift, probably in unison with larger-scale, local cosmic-web structures 
naturally arising in a $\Lambda$CDM cosmological context, as suggested with
dark matter-only simulations \citep{Libeskind12,Libeskind14}.
Furthermore, the physical behaviour of these satellites (distinguishable distributions of specific angular momenta and pericentric distances from those of  non-kinematically-coherent ones) points to a statistically  distinguishable fate once bound to their host.

This paper presents necessary work involved in disentangling the generic consequences of physical laws acting on satellite systems within a $\Lambda$CDM context.
Specifically,
 here we have highlighted  the generic  role long-term angular momentum conservation and orbital pole clustering   plays at setting KPPs. How this clustering might  have been  set at high redshift  is analyzed in a forthcoming paper (G\'amez-Mar\'{i}n  et al., in prep.).

\section*{Acknowledgements}
We thank the anonymous referee for useful comments.
This work was supported through  MINECO/FEDER (Spain)  AYA2012-31101,  AYA2015-63810-P and MICIIN/FEDER (Spain) PGC2018-094975-C21 grants.
ISS acknowledges support by the European Research Council (ERC) through Advanced Investigator grant to C.S. Frenk, DMIDAS (GA
786910).
MGM thanks MINECO/FEDER funding through a FPI fellowship associated to this grant.
PBT acknowledges partial funding by Fondecyt 1200703/2020 (ANID) and CATA-Basal-FB210003 project.
MAGF acknowledges financial support from the Spanish Ministry of Science and Innovation through the project PID2020-114581GB-C22.
VRP thanks the Comunidad de Madrid, Consejer\'ia de Ciencia, Universidades e Innovaci\'on,  for funding him through  contract number PEJ-2019-TIC-15074.  
This work used the  Geryon cluster (Pontificia Universidad de Chile). 
We used a version of Aq-C-5 that  is part of the CIELO Project run in Marenostrum (Barcelona Supercomputer Centre), the NLHPC (funded by ECM-02) and Ladgerda cluster (Fondecyt 12000703).
This project has received funding from the European Union’s Horizon 2020 Research and Innovation Programme under the Marie Skłodowska-Curie grant agreement No 734374-LACEGAL. 


\bibliography{archive_planes}{}

\begin{thebibliography}{}
\expandafter\ifx\csname natexlab\endcsname\relax\def\natexlab#1{#1}\fi
\providecommand{\url}[1]{\href{#1}{#1}}
\providecommand{\dodoi}[1]{doi:~\href{http://doi.org/#1}{\nolinkurl{#1}}}
\providecommand{\doeprint}[1]{\href{http://ascl.net/#1}{\nolinkurl{http://ascl.net/#1}}}
\providecommand{\doarXiv}[1]{\href{https://arxiv.org/abs/#1}{\nolinkurl{https://arxiv.org/abs/#1}}}

\bibitem[{{Ahmed} {et~al.}(2017){Ahmed}, {Brooks}, \& {Christensen}}]{Ahmed17}
{Ahmed}, S.~H., {Brooks}, A.~M., \& {Christensen}, C.~R. 2017, \mnras, 466,
  3119, \dodoi{10.1093/mnras/stw3271}

\bibitem[{{Bahl} \& {Baumgardt}(2014)}]{Bahl14}
{Bahl}, H., \& {Baumgardt}, H. 2014, \mnras, 438, 2916,
  \dodoi{10.1093/mnras/stt2399}

\bibitem[{{Buck} {et~al.}(2016){Buck}, {Dutton}, \& {Macci{\`o}}}]{Buck16}
{Buck}, T., {Dutton}, A.~A., \& {Macci{\`o}}, A.~V. 2016, \mnras, 460, 4348,
  \dodoi{10.1093/mnras/stw1232}

\bibitem[{{Buck} {et~al.}(2015){Buck}, {Macci{\`o}}, \& {Dutton}}]{Buck15}
{Buck}, T., {Macci{\`o}}, A.~V., \& {Dutton}, A.~A. 2015, \apj, 809, 49,
  \dodoi{10.1088/0004-637X/809/1/49}

\bibitem[{{Cautun} {et~al.}(2015){Cautun}, {Bose}, {Frenk}, {Guo}, {Han},
  {Hellwing}, {Sawala}, \& {Wang}}]{Cautun15}
{Cautun}, M., {Bose}, S., {Frenk}, C.~S., {et~al.} 2015, \mnras, 452, 3838,
  \dodoi{10.1093/mnras/stv1557}

\bibitem[{{Chiboucas} {et~al.}(2013){Chiboucas}, {Jacobs}, {Tully}, \&
  {Karachentsev}}]{Chiboucas2013}
{Chiboucas}, K., {Jacobs}, B.~A., {Tully}, R.~B., \& {Karachentsev}, I.~D.
  2013, \aj, 146, 126, \dodoi{10.1088/0004-6256/146/5/126}

\bibitem[{{Collins} {et~al.}(2015){Collins}, {Martin}, {Rich}, {Ibata},
  {Chapman}, {McConnachie}, {Ferguson}, {Irwin}, \& {Lewis}}]{Collins15}
{Collins}, M. L.~M., {Martin}, N.~F., {Rich}, R.~M., {et~al.} 2015, \apj, 799,
  L13, \dodoi{10.1088/2041-8205/799/1/L13}

\bibitem[{{Conn} {et~al.}(2013){Conn}, {Lewis}, {Ibata}, {Parker}, {Zucker},
  {McConnachie}, {Martin}, {Valls-Gabaud}, {Tanvir}, {Irwin}, {Ferguson}, \&
  {Chapman}}]{Conn13}
{Conn}, A.~R., {Lewis}, G.~F., {Ibata}, R.~A., {et~al.} 2013, \apj, 766, 120,
  \dodoi{10.1088/0004-637X/766/2/120}

\bibitem[{Cram\'er(1999)}]{Cramer}
Cram\'er, H. 1999, Mathematical Methods of Statistics (PMS-9) (Princeton
  University Press).
\newblock \url{http://www.jstor.org/stable/j.ctt1bpm9r4}

\bibitem[{{Dom{\'e}nech-Moral} {et~al.}(2012){Dom{\'e}nech-Moral},
  {Mart{\'{\i}}nez-Serrano}, {Dom{\'{\i}}nguez-Tenreiro}, \&
  {Serna}}]{domenech12}
{Dom{\'e}nech-Moral}, M., {Mart{\'{\i}}nez-Serrano}, F.~J.,
  {Dom{\'{\i}}nguez-Tenreiro}, R., \& {Serna}, A. 2012, \mnras, 421, 2510,
  \dodoi{10.1111/j.1365-2966.2012.20534.x}

\bibitem[{{Drlica-Wagner} {et~al.}(2015){Drlica-Wagner}, {Bechtol}, {Rykoff},
  {Luque}, {Queiroz}, {Mao}, {Wechsler}, {Simon}, {Santiago}, {Yanny},
  {Balbinot}, {Dodelson}, {Fausti Neto}, {James}, {Li}, {Maia}, {Marshall},
  {Pieres}, {Stringer}, {Walker}, {Abbott}, {Abdalla}, {Allam},
  {Benoit-L{\'e}vy}, {Bernstein}, {Bertin}, {Brooks}, {Buckley-Geer}, {Burke},
  {Carnero Rosell}, {Carrasco Kind}, {Carretero}, {Crocce}, {da Costa},
  {Desai}, {Diehl}, {Dietrich}, {Doel}, {Eifler}, {Evrard}, {Finley},
  {Flaugher}, {Fosalba}, {Frieman}, {Gaztanaga}, {Gerdes}, {Gruen}, {Gruendl},
  {Gutierrez}, {Honscheid}, {Kuehn}, {Kuropatkin}, {Lahav}, {Martini},
  {Miquel}, {Nord}, {Ogando}, {Plazas}, {Reil}, {Roodman}, {Sako}, {Sanchez},
  {Scarpine}, {Schubnell}, {Sevilla-Noarbe}, {Smith}, {Soares-Santos},
  {Sobreira}, {Suchyta}, {Swanson}, {Tarle}, {Tucker}, {Vikram}, {Wester},
  {Zhang}, {Zuntz}, \& {DES Collaboration}}]{Drlica-Wagner2015}
{Drlica-Wagner}, A., {Bechtol}, K., {Rykoff}, E.~S., {et~al.} 2015, \apj, 813,
  109, \dodoi{10.1088/0004-637X/813/2/109}

\bibitem[{{Fernando} {et~al.}(2017){Fernando}, {Arias}, {Guglielmo}, {Lewis},
  {Ibata}, \& {Power}}]{Fernando17}
{Fernando}, N., {Arias}, V., {Guglielmo}, M., {et~al.} 2017, \mnras, 465, 641,
  \dodoi{10.1093/mnras/stw2694}

\bibitem[{{Forero-Romero} \& {Arias}(2018)}]{ForeroRomero2018}
{Forero-Romero}, J.~E., \& {Arias}, V. 2018, \mnras, 478, 5533,
  \dodoi{10.1093/mnras/sty1349}

\bibitem[{{Fritz} {et~al.}(2018){Fritz}, {Battaglia}, {Pawlowski},
  {Kallivayalil}, {van der Marel}, {Sohn}, {Brook}, \& {Besla}}]{Fritz18}
{Fritz}, T.~K., {Battaglia}, G., {Pawlowski}, M.~S., {et~al.} 2018, \aap, 619,
  A103, \dodoi{10.1051/0004-6361/201833343}

\bibitem[{{Gaia Collaboration} {et~al.}(2018){Gaia Collaboration}, {Helmi},
  {van Leeuwen}, {McMillan}, {Massari}, {Antoja}, {Robin}, {Lindegren},
  {Bastian}, {Arenou}, \& et~al.}]{GaiaHelmi18}
{Gaia Collaboration}, {Helmi}, A., {van Leeuwen}, F., {et~al.} 2018, \aap, 616,
  A12, \dodoi{10.1051/0004-6361/201832698}

\bibitem[{{Garaldi} {et~al.}(2018){Garaldi}, {Romano-D{\'\i}az},
  {Borzyszkowski}, \& {Porciani}}]{Garaldi18}
{Garaldi}, E., {Romano-D{\'\i}az}, E., {Borzyszkowski}, M., \& {Porciani}, C.
  2018, \mnras, 473, 2234, \dodoi{10.1093/mnras/stx2489}

\bibitem[{{Garavito-Camargo} {et~al.}(2021){Garavito-Camargo}, {Patel},
  {Besla}, {Price-Whelan}, {G{\'o}mez}, {Laporte}, \&
  {Johnston}}]{GaravitoCamargo2021}
{Garavito-Camargo}, N., {Patel}, E., {Besla}, G., {et~al.} 2021, \apj, 923,
  140, \dodoi{10.3847/1538-4357/ac2c05}

\bibitem[{{Gillet} {et~al.}(2015){Gillet}, {Ocvirk}, {Aubert}, {Knebe},
  {Libeskind}, {Yepes}, {Gottl{\"o}ber}, \& {Hoffman}}]{Gillet15}
{Gillet}, N., {Ocvirk}, P., {Aubert}, D., {et~al.} 2015, \apj, 800, 34,
  \dodoi{10.1088/0004-637X/800/1/34}

\bibitem[{{Ibata} {et~al.}(2014){Ibata}, {Ibata}, {Lewis}, {Martin}, {Conn},
  {Elahi}, {Arias}, \& {Fernando}}]{Ibata2014}
{Ibata}, R.~A., {Ibata}, N.~G., {Lewis}, G.~F., {et~al.} 2014, \apjl, 784, L6,
  \dodoi{10.1088/2041-8205/784/1/L6}

\bibitem[{{Ibata} {et~al.}(2013){Ibata}, {Lewis}, {Conn}, {Irwin},
  {McConnachie}, {Chapman}, {Collins}, {Fardal}, {Ferguson}, {Ibata}, {Mackey},
  {Martin}, {Navarro}, {Rich}, {Valls-Gabaud}, \& {Widrow}}]{Ibata13}
{Ibata}, R.~A., {Lewis}, G.~F., {Conn}, A.~R., {et~al.} 2013, \nat, 493, 62,
  \dodoi{10.1038/nature11717}

\bibitem[{{Koch} \& {Grebel}(2006)}]{Koch06}
{Koch}, A., \& {Grebel}, E.~K. 2006, \aj, 131, 1405, \dodoi{10.1086/499534}

\bibitem[{{Kroupa} {et~al.}(2005){Kroupa}, {Theis}, \& {Boily}}]{Kroupa05}
{Kroupa}, P., {Theis}, C., \& {Boily}, C.~M. 2005, \aap, 431, 517,
  \dodoi{10.1051/0004-6361:20041122}

\bibitem[{{Kunkel} \& {Demers}(1976)}]{Kunkel76}
{Kunkel}, W.~E., \& {Demers}, S. 1976, in The Galaxy and the Local Group, Vol.
  182, 241

\bibitem[{{Libeskind} {et~al.}(2005){Libeskind}, {Frenk}, {Cole}, {Helly},
  {Jenkins}, {Navarro}, \& {Power}}]{Libeskind05}
{Libeskind}, N.~I., {Frenk}, C.~S., {Cole}, S., {et~al.} 2005, \mnras, 363,
  146, \dodoi{10.1111/j.1365-2966.2005.09425.x}

\bibitem[{{Libeskind} {et~al.}(2009){Libeskind}, {Frenk}, {Cole}, {Jenkins}, \&
  {Helly}}]{Libeskind09}
{Libeskind}, N.~I., {Frenk}, C.~S., {Cole}, S., {Jenkins}, A., \& {Helly},
  J.~C. 2009, \mnras, 399, 550, \dodoi{10.1111/j.1365-2966.2009.15315.x}

\bibitem[{{Libeskind} {et~al.}(2012){Libeskind}, {Hoffman}, {Knebe},
  {Steinmetz}, {Gottl{\"o}ber}, {Metuki}, \& {Yepes}}]{Libeskind12}
{Libeskind}, N.~I., {Hoffman}, Y., {Knebe}, A., {et~al.} 2012, \mnras, 421,
  L137, \dodoi{10.1111/j.1745-3933.2012.01222.x}

\bibitem[{{Libeskind} {et~al.}(2014){Libeskind}, {Knebe}, {Hoffman}, \&
  {Gottl{\"o}ber}}]{Libeskind14}
{Libeskind}, N.~I., {Knebe}, A., {Hoffman}, Y., \& {Gottl{\"o}ber}, S. 2014,
  \mnras, 443, 1274, \dodoi{10.1093/mnras/stu1216}

\bibitem[{{Lipnicky} \& {Chakrabarti}(2017)}]{Lipnicky17}
{Lipnicky}, A., \& {Chakrabarti}, S. 2017, \mnras, 468, 1671,
  \dodoi{10.1093/mnras/stx286}

\bibitem[{{L{\'o}pez} {et~al.}(2019){L{\'o}pez}, {Merch{\'a}n}, \&
  {Paz}}]{Lopez2019}
{L{\'o}pez}, P., {Merch{\'a}n}, M.~E., \& {Paz}, D.~J. 2019, \mnras, 485, 5244,
  \dodoi{10.1093/mnras/stz762}

\bibitem[{{Lovell} {et~al.}(2011){Lovell}, {Eke}, {Frenk}, \&
  {Jenkins}}]{Lovell2011}
{Lovell}, M.~R., {Eke}, V.~R., {Frenk}, C.~S., \& {Jenkins}, A. 2011, \mnras,
  413, 3013, \dodoi{10.1111/j.1365-2966.2011.18377.x}

\bibitem[{{Lynden-Bell}(1976)}]{Lynden76}
{Lynden-Bell}, D. 1976, \mnras, 174, 695, \dodoi{10.1093/mnras/174.3.695}

\bibitem[{{Maji} {et~al.}(2017){Maji}, {Zhu}, {Marinacci}, \& {Li}}]{Maji17b}
{Maji}, M., {Zhu}, Q., {Marinacci}, F., \& {Li}, Y. 2017, \apj, 843, 62,
  \dodoi{10.3847/1538-4357/aa72f5}

\bibitem[{{Mart{\'\i}nez-Delgado} {et~al.}(2021){Mart{\'\i}nez-Delgado},
  {Makarov}, {Javanmardi}, {Pawlowski}, {Makarova}, {Donatiello}, {Lang},
  {Rom{\'a}n}, {Vivas}, \& {Carballo-Bello}}]{MtnezDelgado2021}
{Mart{\'\i}nez-Delgado}, D., {Makarov}, D., {Javanmardi}, B., {et~al.} 2021,
  \aap, 652, A48, \dodoi{10.1051/0004-6361/202141242}

\bibitem[{{Mart{\'\i}nez-Serrano} {et~al.}(2008){Mart{\'\i}nez-Serrano},
  {Serna}, {Dom{\'\i}nguez- Tenreiro}, \& {Moll{\'a}}}]{MartinezSerrano08}
{Mart{\'\i}nez-Serrano}, F.~J., {Serna}, A., {Dom{\'\i}nguez- Tenreiro}, R., \&
  {Moll{\'a}}, M. 2008, \mnras, 388, 39,
  \dodoi{10.1111/j.1365-2966.2008.13383.x}

\bibitem[{{McConnachie} \& {Irwin}(2006)}]{McConnachie06}
{McConnachie}, A.~W., \& {Irwin}, M.~J. 2006, \mnras, 365, 902,
  \dodoi{10.1111/j.1365-2966.2005.09771.x}

\bibitem[{{Metz} {et~al.}(2007){Metz}, {Kroupa}, \& {Jerjen}}]{Metz07}
{Metz}, M., {Kroupa}, P., \& {Jerjen}, H. 2007, \mnras, 374, 1125,
  \dodoi{10.1111/j.1365-2966.2006.11228.x}

\bibitem[{{Metz} {et~al.}(2008){Metz}, {Kroupa}, \& {Libeskind}}]{Metz08}
{Metz}, M., {Kroupa}, P., \& {Libeskind}, N.~I. 2008, \apj, 680, 287,
  \dodoi{10.1086/587833}

\bibitem[{{M{\"u}ller} {et~al.}(2016){M{\"u}ller}, {Jerjen}, {Pawlowski}, \&
  {Binggeli}}]{Muller2016}
{M{\"u}ller}, O., {Jerjen}, H., {Pawlowski}, M.~S., \& {Binggeli}, B. 2016,
  \aap, 595, A119, \dodoi{10.1051/0004-6361/201629298}

\bibitem[{{M{\"u}ller} {et~al.}(2018){M{\"u}ller}, {Pawlowski}, {Jerjen}, \&
  {Lelli}}]{Muller2018}
{M{\"u}ller}, O., {Pawlowski}, M.~S., {Jerjen}, H., \& {Lelli}, F. 2018,
  Science, 359, 534, \dodoi{10.1126/science.aao1858}

\bibitem[{{M{\"u}ller} {et~al.}(2017){M{\"u}ller}, {Scalera}, {Binggeli}, \&
  {Jerjen}}]{Muller2017}
{M{\"u}ller}, O., {Scalera}, R., {Binggeli}, B., \& {Jerjen}, H. 2017, \aap,
  602, A119, \dodoi{10.1051/0004-6361/201730434}

\bibitem[{{M{\"u}ller} {et~al.}(2021){M{\"u}ller}, {Pawlowski}, {Lelli},
  {Fahrion}, {Rejkuba}, {Hilker}, {Kanehisa}, {Libeskind}, \&
  {Jerjen}}]{Muller2021}
{M{\"u}ller}, O., {Pawlowski}, M.~S., {Lelli}, F., {et~al.} 2021, \aap, 645,
  L5, \dodoi{10.1051/0004-6361/202039973}

\bibitem[{{Pawlowski} \& {Kroupa}(2013)}]{Pawlowski13b}
{Pawlowski}, M.~S., \& {Kroupa}, P. 2013, \mnras, 435, 2116,
  \dodoi{10.1093/mnras/stt1429}

\bibitem[{{Pawlowski} \& {Kroupa}(2020)}]{Pawlowski2020}
---. 2020, \mnras, 491, 3042, \dodoi{10.1093/mnras/stz3163}

\bibitem[{{Pawlowski} {et~al.}(2013){Pawlowski}, {Kroupa}, \&
  {Jerjen}}]{Pawlowski13}
{Pawlowski}, M.~S., {Kroupa}, P., \& {Jerjen}, H. 2013, \mnras, 435, 1928,
  \dodoi{10.1093/mnras/stt1384}

\bibitem[{{Pawlowski} {et~al.}(2012){Pawlowski}, {Pflamm-Altenburg}, \&
  {Kroupa}}]{Pawlowski12}
{Pawlowski}, M.~S., {Pflamm-Altenburg}, J., \& {Kroupa}, P. 2012, \mnras, 423,
  1109, \dodoi{10.1111/j.1365-2966.2012.20937.x}

\bibitem[{{Pedrosa} \& {Tissera}(2015)}]{Pedrosa15}
{Pedrosa}, S.~E., \& {Tissera}, P.~B. 2015, \aap, 584, A43,
  \dodoi{10.1051/0004-6361/201526440}

\bibitem[{{Samuel} {et~al.}(2021){Samuel}, {Wetzel}, {Chapman}, {Tollerud},
  {Hopkins}, {Boylan-Kolchin}, {Bailin}, \& {Faucher-Gigu{\`e}re}}]{Samuel2021}
{Samuel}, J., {Wetzel}, A., {Chapman}, S., {et~al.} 2021, \mnras, 504, 1379,
  \dodoi{10.1093/mnras/stab955}

\bibitem[{{Santos-Santos} {et~al.}(2020{\natexlab{a}}){Santos-Santos},
  {Dom{\'\i}nguez-Tenreiro}, {Artal}, {Pedrosa}, {Bignone},
  {Mart{\'\i}nez-Serrano}, {G{\'o}mez-Flechoso}, {Tissera}, \&
  {Serna}}]{SantosSantos2020II}
{Santos-Santos}, I., {Dom{\'\i}nguez-Tenreiro}, R., {Artal}, H., {et~al.}
  2020{\natexlab{a}}, \apj, 897, 71, \dodoi{10.3847/1538-4357/ab7f29}

\bibitem[{{Santos-Santos} {et~al.}(2020{\natexlab{b}}){Santos-Santos},
  {Dom{\'\i}nguez-Tenreiro}, \& {Pawlowski}}]{SantosSantos2020I}
{Santos-Santos}, I.~M., {Dom{\'\i}nguez-Tenreiro}, R., \& {Pawlowski}, M.~S.
  2020{\natexlab{b}}, \mnras, 499, 3755, \dodoi{10.1093/mnras/staa3130}

\bibitem[{{Sawala} {et~al.}(2022){Sawala}, {Cautun}, {Frenk}, {Helly},
  {Jasche}, {Jenkins}, {Johansson}, {Lavaux}, {McAlpine}, \&
  {Schaller}}]{Sawala2022}
{Sawala}, T., {Cautun}, M., {Frenk}, C.~S., {et~al.} 2022, arXiv e-prints,
  arXiv:2205.02860.
\newblock \doarXiv{2205.02860}

\bibitem[{{Scannapieco} {et~al.}(2005){Scannapieco}, {Tissera}, {White}, \&
  {Springel}}]{Scannapieco05}
{Scannapieco}, C., {Tissera}, P.~B., {White}, S.~D.~M., \& {Springel}, V. 2005,
  \mnras, 364, 552, \dodoi{10.1111/j.1365-2966.2005.09574.x}

\bibitem[{{Scannapieco} {et~al.}(2006){Scannapieco}, {Tissera}, {White}, \&
  {Springel}}]{Scannapieco06}
---. 2006, \mnras, 371, 1125, \dodoi{10.1111/j.1365-2966.2006.10785.x}

\bibitem[{{Shao} {et~al.}(2019){Shao}, {Cautun}, \& {Frenk}}]{Shao19}
{Shao}, S., {Cautun}, M., \& {Frenk}, C.~S. 2019, \mnras, 1692,
  \dodoi{10.1093/mnras/stz1741}

\bibitem[{{Sohn} {et~al.}(2020){Sohn}, {Patel}, {Fardal}, {Besla}, {van der
  Marel}, {Geha}, \& {Guhathakurta}}]{Sohn2020}
{Sohn}, S.~T., {Patel}, E., {Fardal}, M.~A., {et~al.} 2020, \apj, 901, 43,
  \dodoi{10.3847/1538-4357/abaf49}

\bibitem[{{Springel} {et~al.}(2001){Springel}, {Yoshida}, \&
  {White}}]{Springel01}
{Springel}, V., {Yoshida}, N., \& {White}, S. D.~M. 2001, \na, 6, 79,
  \dodoi{10.1016/S1384-1076(01)00042-2}

\bibitem[{{Springel} {et~al.}(2008){Springel}, {Wang}, {Vogelsberger},
  {Ludlow}, {Jenkins}, {Helmi}, {Navarro}, {Frenk}, \& {White}}]{Springel08}
{Springel}, V., {Wang}, J., {Vogelsberger}, M., {et~al.} 2008, \mnras, 391,
  1685, \dodoi{10.1111/j.1365-2966.2008.14066.x}

\bibitem[{{Tully} {et~al.}(2015){Tully}, {Libeskind}, {Karachentsev},
  {Karachentseva}, {Rizzi}, \& {Shaya}}]{Tully2015}
{Tully}, R.~B., {Libeskind}, N.~I., {Karachentsev}, I.~D., {et~al.} 2015,
  \apjl, 802, L25, \dodoi{10.1088/2041-8205/802/2/L25}

\bibitem[{{Wang} {et~al.}(2013){Wang}, {Frenk}, \& {Cooper}}]{Wang13}
{Wang}, J., {Frenk}, C.~S., \& {Cooper}, A.~P. 2013, \mnras, 429, 1502,
  \dodoi{10.1093/mnras/sts442}

\end{thebibliography}
\bibliographystyle{aasjournal}


\appendix
\restartappendixnumbering

\section{The `Scanning of Stacked  Orbital Poles Method' applied to the PDEVA-5004 simulation}
\label{appendix1}

\begin{figure}[H]
\centering
\includegraphics[width=0.7\linewidth]{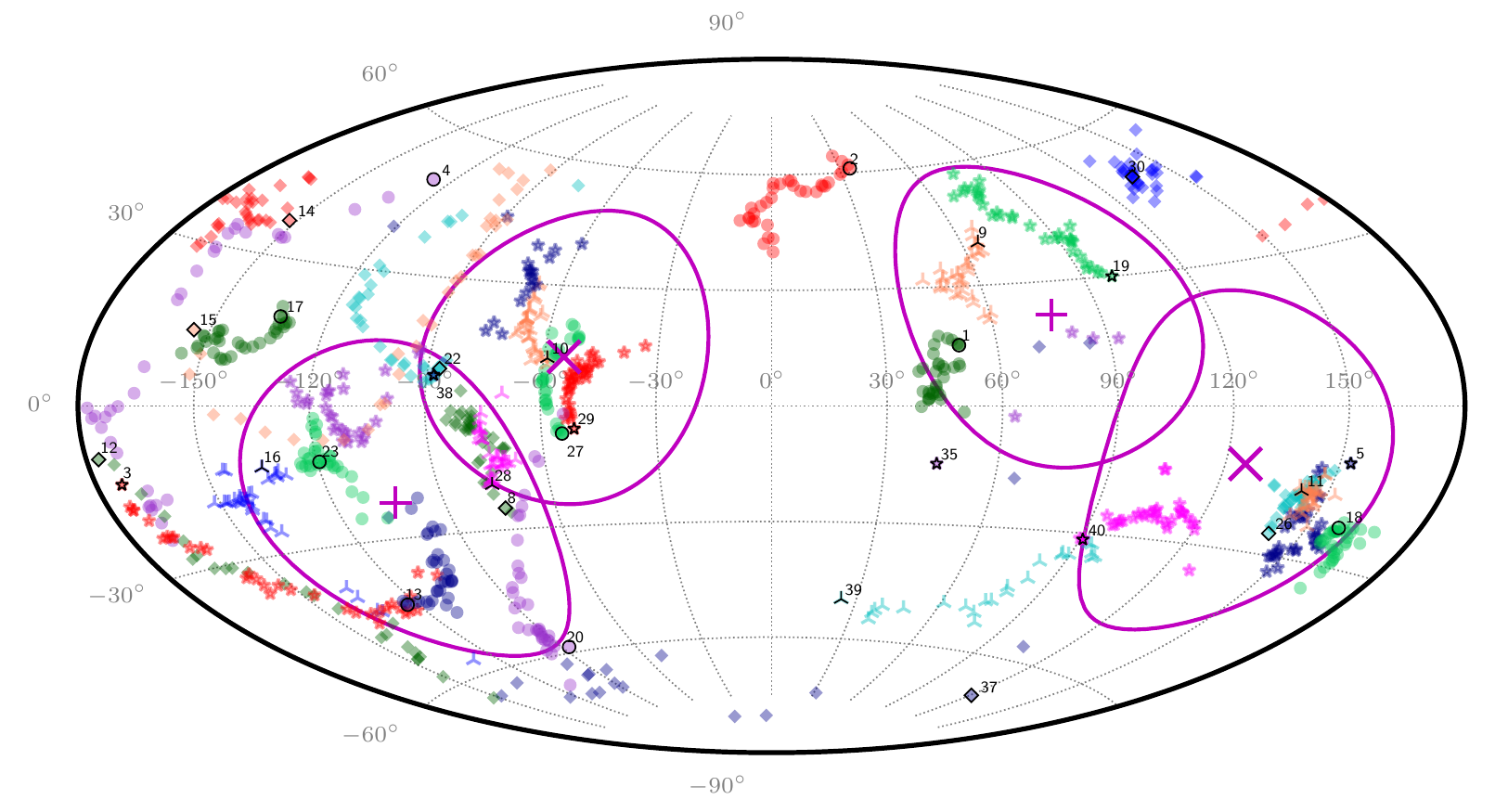}
\caption{Same content as  Figure~\ref{fig:Aitoff_AqC_ALL}A but for the PDEVA-5004 system:
Aitoff projection of the stacked orbital poles for the  whole set of PDEVA-5004 satellites.
Different colors correspond to different satellite tracks. Within a given color, each point corresponds to a simulation output, that is, different values of the universe age, T$_{\rm uni}$, from T$_{\rm vir}$ to T$_{\rm z=0}$.
Two $\vec{J}_{\rm stack}$ axes are identified, marked with large `x' (KPP1) and `+' (KPP2) crosses. The IDs of satellites in each KPP plane can be read in Table \ref{table:summary-discard}.
For more details, see descriptions corresponding to Figure~\ref{fig:Aitoff_AqC_ALL}A.
}
\label{fig:PDEVA-Jsatck}
\end{figure}

\section{The `3$\vec{J}_{\rm orb}$-barycenter Method' to determine the axes of maximum co-orbitation}
\label{appendix2}

This method was described in \citet{SantosSantos2020I} (Paper I) where it was used on MW satellite $z=0$ orbital pole data. We refer the reader to this paper for details.
We briefly describe here its application to simulations, which consist of many timesteps. In this case, the method uses, at each timestep, the directions of the barycenters of all the possible spherical triangles that can be formed from the satellite orbital pole projections on the sphere.
At  each timestep, a direction of maximum satellite co-orbitation is identified by scanning the sphere with an aperture $\Delta_{\rm scan}$ (as in the `Scanning of Stacked Orbital Poles Method') but on these barycenters, instead of the satellite orbital poles. As a result, an independent axis of maximum co-orbitation is found for each timestep. Then, a prescripion to ensure the clustering of conserved satellite orbital poles across time is necessary. We adopt the same protocol as that used in Section~\ref{sec:MemberDet}, but now considering the different axes obtained at each timestep instead of the fixed $\vec{J}_{\rm stack}$. Results for KPP satellite membership using this alternative method are given in Table \ref{table:summary-3JB}.

The differences  regarding KPP member identities in the Aq-C$^\alpha$ and PDEVA-5004 systems  obtained with the ``barycenter" or the ``stacking" method can be read by crossing columns in Tables \ref{table:summary-discard} and \ref{table:summary-3JB}. 
For  Aq-C$^\alpha$ the difference
is of one satellite.
In the case of PDEVA-5004, both methods give exactly the same KPP1 and KPP2 plane members with 10 and 7 satellites respectively.

\begin{table*}
\centering
\scriptsize
\caption{Same as Table \ref{table:summary-discard} for the `3$\vec{J}_{\rm orb}$-barycenter Method'.
}
\begin{tabular}{ l l l }
\toprule
\midrule
   Aq-C$^\alpha$ & PDEVA-5004 (1) & PDEVA-5004 (2) \\
\midrule
    349    &   \red{10}  & \red{13}   \\
\red{353}  &    11     &  19 \\
\red{346}  &    26      & \red{23}  \\
   506     &    \red{27}  & 9  \\
\red{341}  &   \red{29}    & \red{16}  \\
\red{14}   &   5  &  \red{20} \\
\red{21}   &   40   &  1 (4/3) \\
     29    &    \red{8} (4/4) &   \\
    478    &    \red{28}   &   \\
    345    &    18 (5/5) &   \\
    474     &           &   \\
\red{343}(6/6)    &    &  \\
\red{352} (7/6) &     &   \\
\midrule
\bottomrule
\end{tabular}
\label{table:summary-3JB}
\end{table*}

\section{Isotropy tests in the `Scanning of Stacked  Orbital Poles Method'}  
\label{appendix3}

The aim of isotropy tests is to prove that the signal we detect, for example in Figure~\ref{fig:unocos}, does not come from a bias introduced  by  the method we use.
To this end, the `Scanning of stacked orbital poles' method is applied to a number N$_{\rm random}$ of isotropized distributions of satellite orbital poles, yielding,
at each realization $i$, the direction of the axis of maximum satellite co-orbitation corresponding to the $i$-th random realization, $\vec{J}_{\rm stack,i}$.

When using the `stacking'  method, the isotropization process is somewhat more involved than usual.  
Here, timesteps are not individually considered, as the ``trajectories" of satellite orbital pole projections on the sphere
  (determined by Newton's laws) have to be taken into account.
 For this purpose, we proceeded as follows: 
 
\begin{enumerate}[label=(\roman*),leftmargin=*,noitemsep,nolistsep]
\item We assume that the  degree of orbital pole  conservation of the satellite set to be isotropized
is the same as shown in  Figure~\ref{fig:Aitoff_AqC_ALL}A  for the Aq-C$^\alpha$  set or 
in Figure~\ref{fig:PDEVA-Jsatck} for the orbital poles of the PDEVA-5004 system. 
If N$_{\rm out}$ is the number of stacked together snapshots used to calculate the $\vec{J}_{\rm stack}$ axis in a given time interval, we mark the direction  on the sphere of the N$_{\rm out}$/2-th  snapshot for each of the satellites in the set (i.e., its median), $\vec{m}_{\rm orb, k}$, where k=1, 2, ..., N$_{\rm sat}$. 
\item  We randomly and independently move  each of these $\vec{m}_{\rm orb, k}$ anchor vectors on the sphere. The set of 
``trajectory" points belonging to each satellite is moved on the sphere together with their corresponding anchor direction.
\item The $k$-th satellite's trajectory points are rotated around their corresponding anchor direction $\vec{m}_{\rm orb, k}$ by a random angle. We made a  different random shot   for each satellite trajectory pattern of points. 
\item We then use this set of N$_{\rm sat} \times$  N$_{\rm out}$  points to calculate their $\vec{J}_{\rm stack, i}$ axis, corresponding to the $i$-th realization.
\item The next step is to calculate $f_{\rm i}(\alpha)$ at a given timestep T$_{\rm uni}$, in the $i$-th realization.
To this end, in each realization $i$, we singled out the orbital poles of the $k$= 1,2, ..., N$_{\rm sat}$ satellites 
at this particular T$_{\rm uni}$, $\vec{n}_{\rm orb, k, i}$.
Then, we calculate the angles $\beta_{k,i}$ from the $\vec{J}_{\rm stack, i}$ axis to the $k$-th direction $\vec{n}_{\rm orb, k, i}$ in this particular $i$-th  realization.
We calculate the corresponding fraction of directions with their $\beta_{k,i}$  angles lower than a given value  $\alpha$, $f_{\rm i}(\alpha)$.
\item At fixed timesteps, we plot the  $f_{\rm i}(\alpha)$ fractions  as a function of 1-$\cos(\alpha)$\footnote{Our calculations are made within an hemisphere, therefore 1-$\cos(\alpha)$ runs from 0.0 to 1.0.}.
\item To finish the exercise, we calculate, for each 1-$\cos(\alpha)$ bin,  the mean and dispersion of the $f_{\rm i}(\alpha)$ values. Results are shown as orange lines and shaded bands in Figure~\ref{fig:unocos}.
\end{enumerate}


\end{document}